\documentclass[10pt,twocolumn]{article}
\usepackage[top=1.8cm, bottom=1.7cm, left=1.8cm, right=1.8cm]{geometry}
\usepackage[hyphens]{url}
\usepackage{graphicx}  %
\usepackage{booktabs} %
\usepackage{enumitem}
\usepackage{times}
\usepackage[keeplastbox]{flushend}
\usepackage{csquotes} %
\PassOptionsToPackage{hyphens}{url}
\usepackage{float}
\usepackage{subfigure}
\usepackage{array}
\usepackage{indentfirst}
\usepackage{tcolorbox}
\usepackage[small,bf]{caption}
\usepackage[marginal,hang]{footmisc}
  
\makeatletter
\def\url@leostyle{%
  \@ifundefined{selectfont}{\def\UrlFont{\small}}%
  {\def\UrlFont{}}%
}
\makeatother
\renewenvironment{thebibliography}[1]{
  \begin{oldthebibliography}{#1}
    \setlength{\itemsep}{0.1em}
    \setlength{\parskip}{0.1em}
}
{
  \end{oldthebibliography}
}

\renewcommand{\footnoterule}{%
  \kern -3pt
  \hrule width 1in 
  \kern 2pt
}

\renewcommand{\footnotesize}{\fontsize{8}{9}\selectfont}

\usepackage[compact]{titlesec}
\titlespacing*{\section}{0pt}{*2}{2pt} 
\titlespacing{\subsection}{0pt}{*2}{2pt}

\newcommand{\reduce}{\vspace{-3pt}}
\newcommand{\reduceA}{\vspace{-2pt}}
\newcommand{\reduceB}{\vspace{-2pt}}

\newcommand{\descr}[1]{\smallskip\noindent\textbf{#1}}

\definecolor{darkgreen}{RGB}{130,0,0}
\definecolor{darkblue}{RGB}{57,79,99}
\usepackage[bookmarks=false, colorlinks=true, plainpages = false,  linkcolor = darkgreen,   citecolor = darkgreen, urlcolor = darkblue, filecolor = blue]{hyperref}
\PassOptionsToPackage{hyphens}{url}\usepackage{hyperref}

\usepackage{ifthen}
\newif\ifshort
 \shortfalse %

\ifshort
  \newcommand{\isShort}{true}
\else
  \newcommand{\isShort}{false}
\fi
\newcommand{\shortVer}[1]{\ifthenelse{\equal{\isShort}{true}}{{#1}}{}}
\newcommand{\longVer}[1]{\ifthenelse{\equal{\isShort}{false}}{{#1}}{}}

\begin{document}

 \title{\bf ``23andMe confirms: I'm super white''\\Analyzing Twitter Discourse On Genetic Testing}

\author{Alexandros Mittos$^1$, Jeremy Blackburn$^2$, and Emiliano De Cristofaro$^1$\\[1ex]
\normalsize $^1$University College London~~~$^2$University of Alabama at Birmingham\\
\normalsize \{a.mittos, e.decristofaro\}@ucl.ac.uk, blackburn@uab.edu}
\date{}

\maketitle

\thispagestyle{plain}
\pagestyle{plain}

\begin{abstract}
Recent progress in genomics is bringing genetic testing to the masses. Participatory public initiatives are underway to sequence the genome of millions of volunteers, and a new market is booming with a number of companies like 23andMe and AncestryDNA offering affordable tests directly to consumers. Consequently, news, experiences, and views on genetic testing are increasingly shared and discussed online and on social networks like Twitter. In this paper, we present a large-scale analysis of Twitter discourse on genetic testing. We collect 302K tweets from 113K users, posted over 2.5 years, by using thirteen keywords related to genetic testing companies and public initiatives as search keywords. We study both the tweets and the users posting them along several axes, aiming to understand who tweets about genetic testing, what they talk about, and how they use Twitter for that. Among other things, we find that tweets about genetic testing originate from accounts that overall appear to be interested in digital health and technology. Also, marketing efforts as well as announcements, such as the FDA's suspension of 23andMe's health reports, influence the type and the nature of user engagement. \longVer{Finally, we report on users who share screenshots of their results, and raise a few ethical and societal questions as we find evidence of groups associating genetic testing to racist ideologies.} 
\end{abstract}

\section{Introduction}\label{sec:intro}
\longVer{
In 1990, the Human Genome Project was kicked off with the goal of producing the first complete sequence of a human genome;
at a cost of almost \$3 billion, it was completed 13 years later~\cite{lander2001initial}.
Since then, costs have dropped at a staggering rate: by 2006, high-quality sequencing of a human genome cost \$14 million, and, by 2016, private individuals could have their genomes sequenced for about \$1,500~\cite{genome2017org}.
This rapid progress}
\shortVer{Progress in genomics }is paving the way to {\em personalized medicine}, a concept advocating for diagnosis and treatment to be tailored to patients' genetic features, aiming to make healthcare more preventive and effective~\cite{ashley2016towards}.
It also enables {\em public initiatives} to sequence large numbers of genomes and build large bio-repositories 
for research purposes; for instance, the Precision Medicine research program in US (now called All Of Us)
or the Genomics England project in UK are sequencing the genomes of, respectively, 1M and 100K volunteers.

Moreover, a number of companies have entered the flourishing market of {\em direct-to-consumer} (DTC) genetic testing.
Rather than visiting a clinic, customers purchase a collection kit for a few hundred dollars or less, deposit a saliva sample, and mail it back;
after a few days, they receive a report with information about genetic health risks (e.g., susceptibility to Alzheimer's), wellness  (e.g., lactose intolerance), and/or ancestry and genealogy information.
Today, there are hundreds of DTC companies\longVer{ -- naturally, some more reputable than others~\cite{phillips2016only} --}
including 23andMe (which provides reports on carrier status, health, and ancestry) and AncestryDNA (which focuses on genealogy and ancestry). 
As of Jan 2018, 23andMe has tested 2M and AncestryDNA 5M customers.\footnote{See \url{http://for.tn/2eYVJuD} and \url{http://ancstry.me/2zH3HBN}}

\longVer{Traditionally, health-related issues were communicated to patients primarily by doctors and clinicians---the advent of direct-to-consumer genetic testing changes this substantially.
Individuals can now learn potentially life-changing results with a few clicks of the mouse, without contacting a medical professional.
Also, as results are delivered electronically, they are more easily shared with others.}
Affordable DTC products and participatory sequencing initiatives make genetic testing increasingly more accessible and available to the general population.
Like with other aspects of digital health, this leads to social media attracting discussions, sharing of experiences, and molding of perceptions around genetic testing, thus becoming a key platform for related news and marketing efforts. 
However,\longVer{ while the research community has analyzed in great detail the interlinked relationship between health and social networks such as Twitter,} to the best of our knowledge, genetic testing discourse on social media has not been adequately studied.

To this end, in this paper, we set to address a few open questions: 
1) What are the tweets related to genetic testing really about? 
2) Which accounts are particularly active in tweeting and what do they talk about? 
3) Is the discussion about genetic testing dominated by certain keywords, themes, or companies? 
4) What is the overall sentiment and what topics relate to more negative sentiment?
(We focus on Twitter due to its popularity and the relatively ease of collecting data).

Aiming to answer these questions, we present an exploratory, large-scale analysis of Twitter discourse related to genetic testing.
Starting from 10 keywords related to DTC genetics companies and 3 to genomics initiatives, we search and crawl all available tweets containing these keywords that were posted between Jan 1, 2015 and Jul 31, 2017. 
We collect 302K tweets from 113K users, and analyze them along several axes\shortVer{. After}\longVer{, seeking to understand who tweets about genetic testing, what they talk about, and how they use Twitter for that. Specifically, after}
presenting a general characterization of our dataset, %
we analyze the tweets content-wise, studying the most common hashtags/URLs and measuring sentiment. %
Next, we perform a user-based analysis, looking at the profiles and their location, and assessing whether they are likely to be bots. %
We also select a random sample of 15K users and analyze their latest 1K tweets to study their interests. 
\longVer{Note that, as a substantial chunk of tweets turns out to be about DTC companies 23andMe and AncestryDNA, we present a few case studies focused on them. %
Finally, we examine the most negative tweets in our dataset, finding a number of tweets related to racism and hate-speech, as well as fears of privacy and data misuse, and look for instances of users sharing screenshots of their test results.}

Overall, our study leads to a few interesting observations:
\ifshort
 \begin{enumerate}
\else
 \begin{enumerate}
\fi 
\item Users tweeting about genetic testing seem overall interested in digital health and technology, although the conversation is often dominated by those with a vested interest in its success, e.g., specialist journalists, medical professionals, entrepreneurs, etc.
\item The two most popular DTC companies, 23andMe and AncestryDNA, also generate the most tweets. However, although 23andMe has half the customers, it produces almost 5 times more tweets, which is also due to controversy around their failure to get FDA approval in 2015. 
\item Sentiment around initiatives is positive, with coverage boosted by mainstream news and announcements (e.g.~President Obama's) and neutral for DTC companies, although with a few strongly opinionated users.
\item There is a clear distinction in the marketing efforts undertaken by different companies, which naturally influence the type and the nature of users' engagement; e.g., we find the the promotional hashtag \#sweepstakes in 1 out 8 tweets containing the keyword AncestryDNA. 
\item We find a limited presence of social bots, with some keywords attracting a different degree of automated publishing, as some topics seem to be more popular among individuals than others. 
\longVer{\item We find evidence of groups using genetic testing to push racist and anti-semitic agendas, and of users expressing concerns about privacy and data protection}
\longVer{\item A non-negligible amount of users share and discuss screenshots of their ancestry test results, despite the possible privacy implications.}  %
\end{enumerate}
\shortVer{Note that, in an extended version of this paper~\cite{full}, we also examine the most negative tweets in our dataset, finding a number of tweets related to racism and hate-speech, as well as fears of privacy and data misuse, and look for instances of users sharing screenshots of their test results.}

\longVer{\descr{Paper organization.} The rest of the paper is organized as follows. 
In Section~\ref{sec:related} we present related work focusing on the following three themes: i) user perspectives on genetic testing, ii) health in social networks, and iii) analysis of social media discourse. 
In Section~\ref{sec:dataset} we present our methodology for collecting the dataset. 
In Section~\ref{sec:content-general} we present a general characterization of the tweets in our dataset. 
In Section~\ref{sec:content-analysis} we study the most common hashtags and URLs, and we measure the overall sentiment of tweets. 
In Section~\ref{sec:user-analysis} we study the profiles of people who tweet about genetic testing. 
Finally, in Section~\ref{sec:use-case} we examine several cases where genetic testing is being used to promote racist agendas, instances of people expressing concerns about privacy, and of people sharing screenshots of their ancestry test results online.

}

\section{Related Work}\label{sec:related}

\descr{User perspectives on genetic testing.} 
A few qualitative studies have analyzed users' perspectives on genetic testing.
Goldsmith et al.~\cite{goldsmith2012direct} review 17 studies conducted in 6 different countries, finding that, although participants appear to be interested in the health-related aspects of testing, they also express concerns about privacy and reliability.\longVer{ Covolo et al.~\cite{covolo2015internet} review 118 articles, highlighting that users are drawn to genetic testing by the potential to monitor and improve their health.}
Caulfield et al.~\cite{caulfield2007myriad} analyze the controversy around Myriad Genetics following their attempt to patent the BRCA gene which is associated with predisposition to breast cancer. 

Closer to our work are quantitative studies using social media, however, to the best of our knowledge, the only relevant work is by Chow-White et al.~\cite{chow2017warren}, who look at one week's worth of tweets containing `23andMe', performing a simple sentiment analysis, finding that positive tweets outnumber negative ones and that people tend to be enthusiastic about it. 
Their analysis only studies one company and only sentiment, whereas, we focus on 10 companies and 3 initiatives and conduct deeper statistical, content, and user-based analyses. 
Also, they rely on a much smaller dataset, 2K vs 324K tweets, collected over 1 week vs 2.5 years. 

\descr{Health in social networks.} 
Social networks \longVer{like Twitter }have been used extensively to study health and health-related issues, e.g., to measure and predict depression. 
De Choudhury et al.~\cite{de2013predicting} identify 476 users self-reporting depression, collect their tweets, and study their engagement, emotion, and use of depressive language. 
By comparing to a control group, they extract significant differences, and build a classifier to predict the likelihood of an individual's depression. 
Coppersmith et al.~\cite{coppersmith2014quantifying} study tweets related to various mental disorders, while Paul et al.~\cite{paul2011you} gather public health information from Twitter, discovering statistically significant correlations between Twitter and official health statistics.
Abbar et al.~\cite{abbar2015you} analyze the nutritional behavior of US citizens: they collect 892K tweets by 400K US users using food-related keywords and find that foods match obesity and diabetes statistics, and that Twitter friends tend to share the same preferences in food consumption. 
\longVer{Prasetyo et al.~\cite{prasetyo2015impact} study how social media can effect awareness in health campaigns.
Focusing on the Movember charity campaign, they collect more than 1M tweets, using the keyword `Movember', and uncover correlations between the visitors of the Movember website and popular Twitter users, but none between tweets and donations.}

\descr{Analyzing social media discourse.} 
Cavazos-Rehg et al.~\cite{cavazos2015hey} study drinking behaviors on Twitter: using keywords related to drinking (e.g., drunk, alcohol, wasted), they collect 10M tweets and identify the most common themes related to pro-drinking and anti-drinking behavior. 
Lerman et al.~\cite{lerman2016emotions} conduct an emotion analysis on tweets from Los Angeles: using public demographic data, they find that users with lower income and education levels, and who engage with less diverse social contacts, express more negative emotions, while people with higher income and education levels post more positive messages.
\longVer{Chatzakou et al.~\cite{chatzakou2017measuring} study the GamerGate controversy\footnote{\url{https://en.wikipedia.org/wiki/Gamergate_controversy}} on Twitter, collecting a dataset of tweets containing keywords indicating abusive behavior. 
They compare the characteristics of the related Twitter profiles to a baseline, finding that users tweeting about Gamergate are more technologically savvy and active, and that their tweets are more negative. }%
Burnap et al.~\cite{burnap2014tweeting} study Twitter response to a terrorist attack occurred in Woolwich in 2013. 
Using `Woolwich' as a keyword search, they collect 427K tweets, finding that opinions and emotional factors are predictive of size and survival of information flows. 

\longVer{
\begin{table*}[t]
\footnotesize
\centering
\begin{tabular}{lrrrrrrrrrr}
\toprule
& {\bf Tweets}  & {\bf Users}   & {\bf RTs}   & {\bf Likes}     	& {\bf Official}  	& {\bf Media}  	& {\bf Quotes}   	& {\bf Hashtags}  	& {\bf URLs}  	& {\bf Top 1M}      \\  \midrule
23andMe                               & 132,597       & 64,014        & 72,848      & 149,897       	& 1.31\%      		& 6.14\%    	& 3.49\%    		& 27.23\%     		& 68.68\%     	& 75.40\%         	\\
AncestryDNA                           & 29,071        & 16,905        & 16,266      & 47,249      		& 7.08\%      		& 8.79\%    	& 2.69\%    		& 54.29\%     		& 75.50\%     	& 49.68\%         	\\
Counsyl                               & 3,862         & 1,834         & 2,716       & 4,255       		& 3.49\%      		& 6.98\%    	& 4.64\%    		& 44.01\%     		& 83.94\%     	& 74.97\%         	\\
DNAFit                                & 2,118         & 844           & 1,336       & 2,508       		& 15.34\%       	& 18.74\%     	& 5.37\%    		& 57.22\%     		& 78.94\%     	& 79.18\%         	\\
FamilyTreeDNA                         & 2,794         & 1,205         & 1,196       & 3,111       		& 4.36\%      		& 19.97\%     	& 6.62\%    		& 34.18\%     		& 36.47\%     	& 69.21\%         	\\
FitnessGenes                          & 2,142         & 773           & 908         & 2,809       		& 16.29\%       	& 18.47\%     	& 9.40\%    		& 44.53\%     		& 56.76\%     	& 71.28\%         	\\
MapMyGenome                           & 1,568         & 704           & 4,488       & 3,726       		& 15.30\%       	& 13.13\%     	& 4.99\%    		& 53.63\%     		& 80.35\%     	& 64.30\%         	\\
PathwayGenomics                       & 1,544         & 579           & 1,968       & 2,521       		& 2.13\%      		& 18.51\%     	& 6.11\%    		& 61.01\%     		& 76.55\%     	& 68.12\%         	\\
Ubiome                                & 14,420        & 6,762         & 9,223       & 13,991      		& 2.71\%      		& 4.37\%    	& 2.85\%    		& 27.85\%     		& 73.28\%     	& 64.19\%         	\\
VeritasGenetics                       & 1,292         & 497           & 1,443       & 2,526       		& 6.65\%      		& 17.07\%     	& 17.07\%     		& 46.13\%     		& 58.28\%     	& 71.95\%         	\\  %
Genomics England                      & 7,009         & 1,863         & 19,772      & 18,756      		& 19.68\%       	& 17.80\%     	& 11.58\%     		& 61.19\%     		& 69.18\%     	& 48.82\%         	\\
Personalized Medicine & 20,302        & 4,631         & 19,085      & 15,514      		& --~         		& 6.93\%    	& 7.55\%    		& 99.93\%     		& 87.42\%   	& 71.98\%         	\\
Precision Medicine                    & 83,329        & 13,012        & 118,043     & 128,303       	& --~         		& 8.56\%    	& 10.41\%     		& 99.88\%       	& 83.39\%   	& 77.16\%       	\\ \midrule
{\em Total}                           & 302,048       & 113,624    & 269,292     		& 395,166       	& 2.26\%      	& 7.75\%    		& 5.92\%    		& 56.54\%     	& 74.77\%     & 71.80\%         						\\ 
{\em Baseline}                        & 163,260       & 131,712    & 282,063,006  		& 486,960,753    	& --~         		& 41.20\%     	& 12.07\%     & 23.48\%       & 45.49\%     & 89.57\%  	\\  \bottomrule
\end{tabular}
\caption{Our keyword dataset, with all tweets from January 1, 2015 to July 31, 2017 containing keywords related to genetic testing.}
\label{tab:keyword_dataset}
\end{table*}}

\shortVer{
\begin{table*}[t]
\footnotesize
\centering
\resizebox{1.3\columnwidth}{!}{%
\renewcommand{\arraystretch}{0.85}
\begin{tabular}{lrrrrrrr}
\toprule
                                      & {\bf Tweets} & {\bf Users}   & {\bf RTs}           & {\bf Likes} 			& {\bf Official}    & {\bf URLs}  & {\bf Top 1M}      	\\  \midrule
23andMe                               & 132,597      & 64,014        & 72,848              & 149,897               & 1.31\%            & 68.68\%     & 75.40\%         	\\
AncestryDNA                           & 29,071       & 16,905        & 16,266              & 47,249                & 7.08\%            & 75.50\%     & 49.68\%         	\\
Counsyl                               & 3,862        & 1,834         & 2,716               & 4,255                 & 3.49\%            & 83.94\%     & 74.97\%         	\\
DNAFit                                & 2,118        & 844           & 1,336               & 2,508                 & 15.34\%           & 78.94\%     & 79.18\%         	\\
FamilyTreeDNA                         & 2,794        & 1,205         & 1,196               & 3,111                 & 4.36\%            & 36.47\%     & 69.21\%         	\\
FitnessGenes                          & 2,142        & 773           & 908                 & 2,809                 & 16.29\%           & 56.76\%     & 71.28\%         	\\
MapMyGenome                           & 1,568        & 704           & 4,488               & 3,726                 & 15.30\%           & 80.35\%     & 64.30\%         	\\
PathwayGenomics      & 1,544        & 579           & 1,968               & 2,521                 & 2.13\%            & 76.55\%     & 68.12\%         	\\
Ubiome                                & 14,420       & 6,762         & 9,223               & 13,991                & 2.71\%            & 73.28\%     & 64.19\%         	\\
VeritasGenetics                       & 1,292        & 497           & 1,443               & 2,526                 & 6.65\%            & 58.28\%     & 71.95\%         	\\  %
Genomics 
England                               & 7,009        & 1,863         & 19,772              & 18,756                & 19.68\%           & 69.18\%     & 48.82\%         	\\
Personalized Medicine 							   & 20,302       & 4,631         & 19,085              & 15,514                & --~              	& 87.42\%     & 71.98\%         	\\
Precision Medicine                    		   & 83,329     & 13,012          & 118,043             & 128,303               & --~              	& 83.39\%     & 77.16\%       		\\  \midrule
{\em Total}                           & 302,048      & 113,624       & 269,292             & 395,166               & 2.26\%            & 74.77\%     & 71.80\%         	\\  %
{\em Baseline}                        & 163,260      & 131,712       & 282,063,006    		& 486,960,753      		& --~              	& 45.49\%     & 89.57\%         	\\  \bottomrule
\end{tabular}
}
\caption{Our keyword dataset, with all tweets from January 1, 2015 to July 31, 2017 containing keywords related to genetic testing.}
\label{tab:keyword_dataset}\reduce
\end{table*}
}

\section{Dataset}\label{sec:dataset}

\longVer{We now present the methodology used to gather the tweets studied in this paper.} 
We build a dataset with tweets containing keywords related to (1) direct-to-consumer (DTC) genetic testing companies, and (2) public genome sequencing initiatives, using these keywords as search queries and crawling all tweets posted from Jan 1, 2015 to Jul 31, 2017 returned as results.

\descr{DTC genetic testing companies.} We start from a list of 36 DTC genetic testing companies compiled by the International Society of Genetic Genealogy\footnote{\url{https://isogg.org/wiki/List_of_DNA_testing_companies}},
which provides a good sample of the DTC ecosystem.
We use each company's name as a search keyword; if the search returns less than 1,000 tweets, we discard it.
In the end, we collect tweets for 10 companies: 23andMe, AncestryDNA, Counsyl, DNAFit, FamilyTreeDNA, FitnessGenes, MapMyGenome, PathwayGenomics, Ubiome, and VeritasGenetics.
We opt for keywords not separated by spaces (e.g., VeritasGenetics) rather than quoted search (e.g., ``Veritas Genetics'') since we notice that companies are primarily discussed via hashtags or mentions, and because Twitter's search engine does not provide exact results with quotes.

\descr{Genomics initiatives.} Besides tweets related to for-profit companies, we also want to study discourse related to public sequencing initiatives and related concepts.
Thus, we select three more keywords: PrecisionMedicine, PersonalizedMedicine, and GenomicsEngland.
Personalized Medicine aims to make diagnosis, treatment, and care of patients tailored and optimized to their specific genetic makeup.
Precision Medicine conveys a similar concept, but also refers to the initiative sequencing the genome of 1M individuals announced by President Obama in 2015\shortVer{.}\longVer{to understand how a person's genetics, environment, and lifestyle can help determine the best approach to prevent or treat disease.}\footnote{\url{https://ghr.nlm.nih.gov/primer/precisionmedicine/initiative}}
Genomics England is a similar UK initiative with 100K volunteers, primarily focusing on cancer and rare disease research.
Once again, we search for keywords not separated by spaces (e.g., PrecisionMedicine) since these concepts are mostly discussed via hashtags and because of the incorrectness of the search engine.

\descr{Crawl.} We use a custom Python script to collect all posted tweets from Jan 1, 2015 to Jul 31, 2017 returned as search results using the 10 DTC keywords and the 3 keywords related to genomics initiatives.
The crawler, run with self-imposed throttling\shortVer{,}\longVer{ to avoid issues for the site operators over} in Fall 2017, collects, for each tweet, its content, the username, date and time, the number of retweets and likes, as well as the URL of the tweet.
It also visits the profile of the users posting each tweet, collecting their location (if any), the number of followers, following, tweets, and likes.
Overall, we collect a total of 191K tweets from 94K users for the 10 DTC companies 
and 111K from 19K users for the 3 initiatives.

\descr{Baseline.} We also collect a set of 163,260 random English tweets, from the same Jan 2015 to Jul 2017 period (approx. 170 per day), which serves as a baseline set for comparisons. 

\descr{Remarks.} The keyword search also returns accounts that match that keyword, e.g., tweets including 23andMe, \#23andMe, or @23andMe, but also those posted by the @23andMe account.
For consistency, we discard the latter, analyzing them separately when relevant. 
Note that our dataset includes tweets from users who discuss their opinions on genetic testing, but also blog posts, ads, news articles, etc.
As our goal is to discover how genetic testing is reflected through the lens of Twitter, we choose not to discard any of the above subsets in an attempt to ``clean'' the dataset, or to focus only on certain kinds of profiles. 
Nevertheless, in Sec.~\ref{sec:user-analysis}, we do shed light on the {\em users} tweeting about genetic testing, as well as those who publish their genetic tests results. 

\section{General Characterization}\label{sec:content-general}
We now present a general characterization of the tweets in our dataset.
Simple statistics of our keyword-based dataset are reported in Table~\ref{tab:keyword_dataset}.
From left to right, the table lists the total number of tweets, unique users, retweets, and likes for each of the 13 keywords and the random baseline. 
We also quantify the percentage of tweets made by the official accounts of each company or initiative, as well as the percentage of \longVer{tweets including media (images and videos), quoted tweets, hashtags, and }%
URLs, and how many of them are in the Alexa Top 1M.

\longVer{
 \begin{figure}[t]
 \centering
 \includegraphics[width=0.34\textwidth]{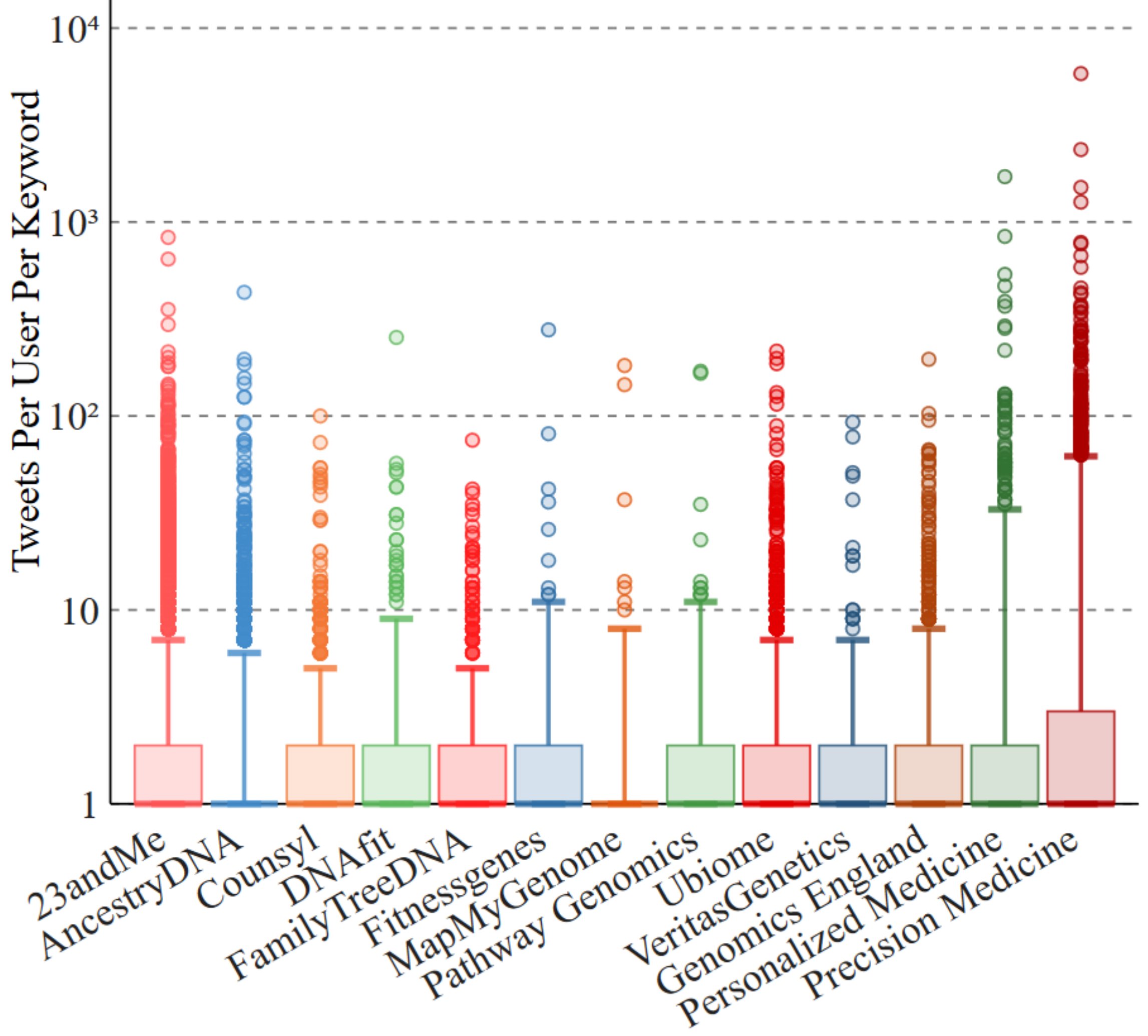}
 \vspace{-0.15cm}
 \caption{Number of tweets per user account.} %
 \label{fig:tweets_per_user_per_keyword}
 \end{figure}
}

\begin{figure}[t]
\centering
\longVer{\includegraphics[width=0.34\textwidth]{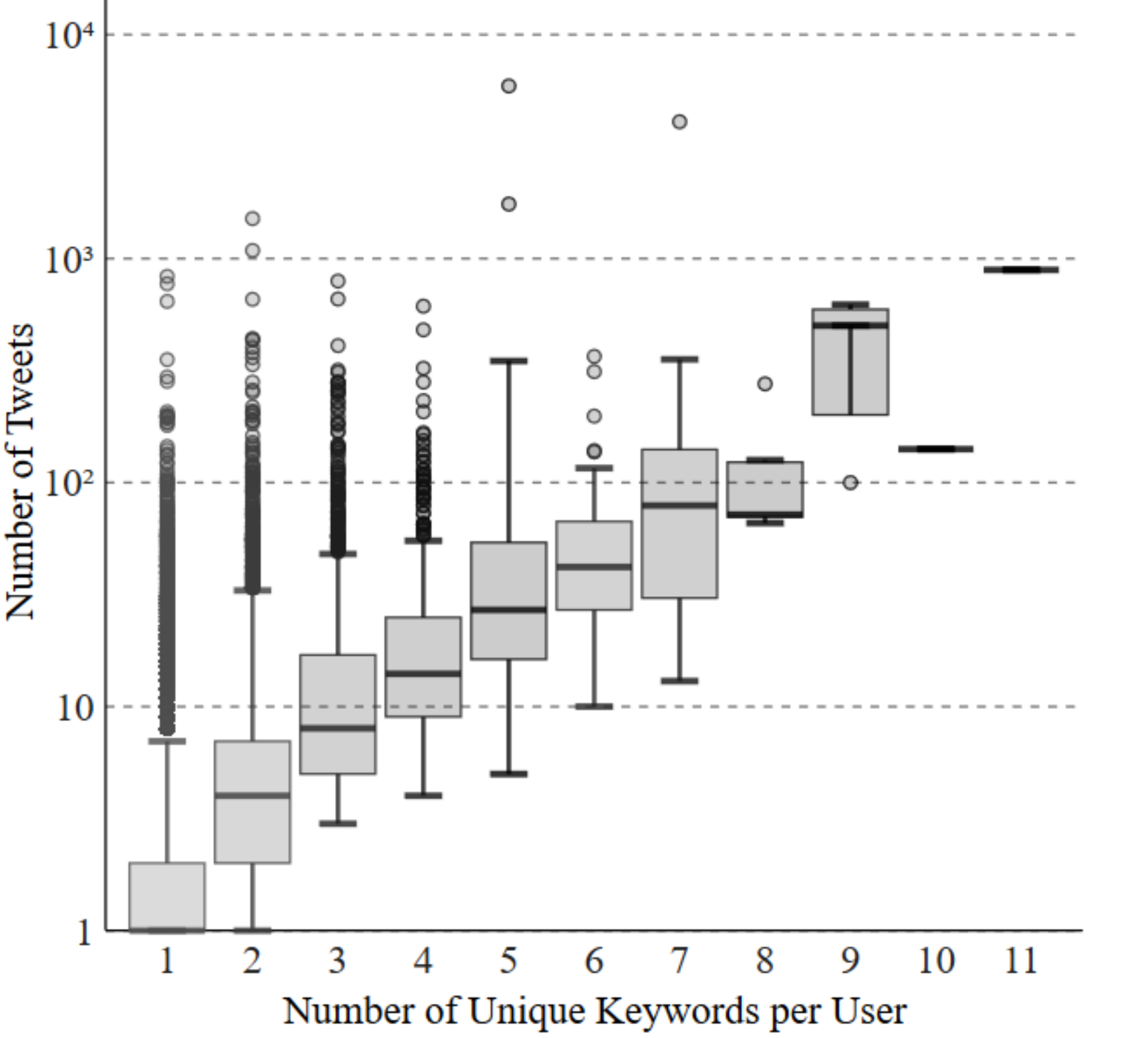}}
\shortVer{\includegraphics[width=0.24\textwidth]{plots/general_char/jeremy.pdf}}
\caption{Number of tweets per user as a function of the number of unique keywords they tweeted about.} %
\label{fig:jeremy}\reduce\reduce
\end{figure}

\descr{DTC vs Initiatives.} Overall, we find differences
between tweets about DTC genetic testing companies and those about genomics initiatives. 
The majority of the latter come from a smaller set of users compared to the former, i.e., a few very dedicated users drive the discussion about genomics initiatives.
\longVer{This is evident from Figure~\ref{fig:tweets_per_user_per_keyword}, which plots the number of tweets per user for each keyword: Personalized Medicine and Precision Medicine have more outliers than most of the DTC genetic companies (although the median for all keywords is 1). }%
We also find these tweets are more likely to contain URLs (87\% and 83\% of tweets, respectively) than most companies, and even more so when compared to the baseline (45\%).
This suggests that tweets about these topics often include links to news and/or other external resources.

Only around 50\% of URLs linked from tweets related to Genomics England or AncestryDNA are in the Alexa top 1M, compared to 60--75\% for other keywords.
For Genomics England, this is due to many URLs pointing to \url{genomicsengland.co.uk} itself.
For AncenstryDNA, whose official site at \url{ancestry.com} \emph{is} in the top 1M, it appears to be due to a very large number of marketing URLs tweeted along with the keyword; we discuss this further later on. 

\descr{Number of tweets.} 
23andMe is by far the most popular keyword, with one order of magnitude more tweets than any other company (130K in total, around 140/day, from 64K distinct users); AncestryDNA is a distant second (30K tweets from 16.9K users).
Given their large customer bases, this should not come as a surprise. 
However, it \emph{is} surprising that 23andMe has 4.6 times as many tweets as AncenstryDNA even though AncestryDNA has over twice the customers as 23andMe.
The least popular companies are MapMyGenome, PathwayGenomics, and VeritasGenetics, with less than 2K tweets each over our 2.5 year collection period.
Among the initiatives, Precision Medicine generates a relative high number of tweets (83K from 13K users), much more so than Personalized Medicine (20K tweets).

\descr{Tweets per user.} 
For each keyword, we also measure the number of tweets per user\shortVer{, although we do not plot it due to space limitations.}\longVer{(see Figure~\ref{fig:tweets_per_user_per_keyword}).}
We find that the median for every keyword is 1; i.e., 50\% of users tweet about a given DTC company or initiative only once. 
However, we do find differences in the outliers for different keywords.
For instance, there are several highly engaged users tweeting about Personalized Medicine and Precision Medicine.
Manual examination of these users indicates that most of them are  medical researchers and companies actively promoting the initiatives as hashtags.
The presence of these heavily ``invested'' users becomes more apparent when we look at the number of tweets as a function of the number of unique keywords a user posts about, as plotted in Figure~\ref{fig:jeremy}:
95\% of them post about only one keyword, and those that post in more than one tend to post \emph{substantially} more tweets about genetic testing in general; in some cases, orders of magnitude more tweets.

\descr{Retweets and Likes.} The total number of retweets and likes per tweet in the baseline is substantially higher than for tweets related to genetic testing due to outliers, i.e., viral tweets or tweets posted by famous accounts (e.g., a tweet by @POTUS44 on January 11, 2017 has 875,844 retweets and 1,862,249 likes).
However, the median for retweets and likes in the baseline dataset mirrors that of tweets in our keywords dataset, with values between 0 and 1. 

Note that, although the number of retweets and likes per tweet could be influenced by how old the tweets are, this is not really the case in our dataset.
We collect tweets posted up to July 2017 starting in late-August 2017, allowing ample time for retweets and likes to occur, considering that previous work~\cite{haewoon2010twitter} indicates that 75\% of retweets happen within 24 hours and 85\% happen within a month.

\descr{Official accounts tweets.} We also look at the tweets including a given keyword (e.g., Ubiome) made by the corresponding official account (e.g., @Ubiome).
There are no official accounts for Personalized and Precision Medicine, however, the Precision Medicine initiative is now called All Of Us and has a Twitter account (created in February 2017) that has posted only a few tweets (224 as of April 4, 2018), so we do not consider it. %
Tweets made by the official accounts of most companies including the name of the company as keyword is unsurprisingly very (e.g., 1\% for 23andMe). 
However, it is higher for others (e.g., 15\% for DNAfit, Fitnessgenes, and MapMyGenome), due to the fact that these companies actually add their names in their tweets as a hashtag (e.g., \#AncestryDNA).
\shortVer{(In fact, we find that hashtags are used quite predominantly for several DTC keywords, in some cases 40\% of tweets have hashtags vs 23\% for baseline tweets.)}

\longVer{
\descr{Hashtags, media, and quotes.} Table~\ref{tab:keyword_dataset} shows that around a quarter of 23andMe's and Ubiome's tweets have hashtags (slightly more than 23\% for the baseline); for most other keywords, it is above 40\%.
For Personalized and Precision Medicine, we find that almost all tweets have the keyword in the form of hashtag (99\%).
For Genomics England, this only happens 61\% of the time, since a lot of tweets include @GenomicsEngland.
We perform a more detailed hashtag analysis in the relevant section. 

We then find that the percentage of tweets with media vary from 4\% in Ubiome to almost 20\% in FamilyTreeDNA. 
Anecdotally, we see that images often contain text, i.e., are used to overcome the character limit and comment on issues related to the company (e.g.,~\cite{ftdna}). %
We also look at ``quotes'', i.e., tweets including the URL of another tweet: for most keywords, percentages are lower than the baseline, except for VeritasGenetics (mostly due to the official account),
although less so for the initiatives.
Possibly, users tweeting about genomics initiatives tend to be discuss more with each other, by commenting on relevant tweets. 
}

\longVer{
\begin{figure}
\centering
\subfigure[23andMe and AncestryDNA]{\includegraphics[width=0.2435\textwidth]{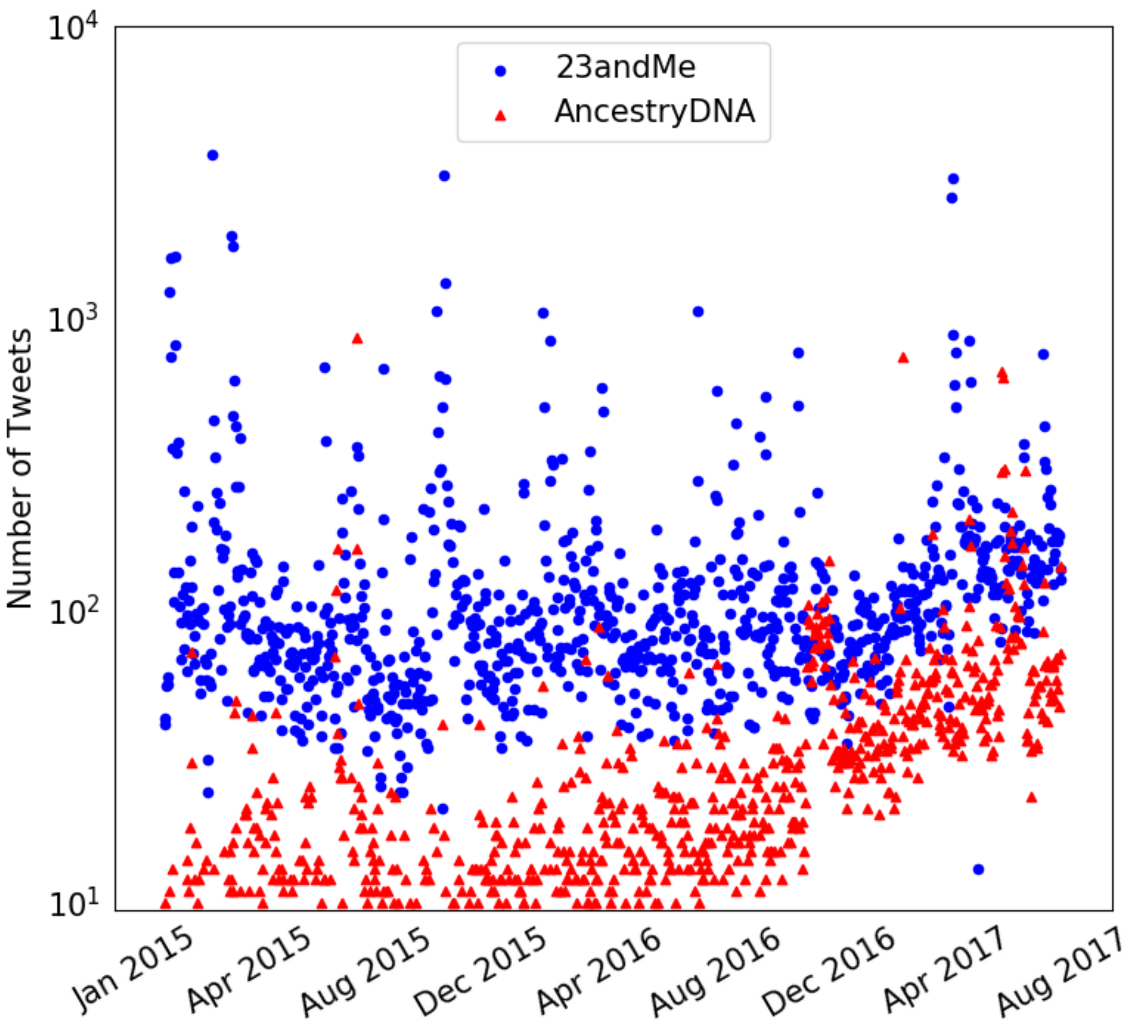}\label{fig:temp-dtc}}
\hspace*{-0.2cm}
\subfigure[Personalized/Precision Medicine]{\includegraphics[width=0.2435\textwidth]{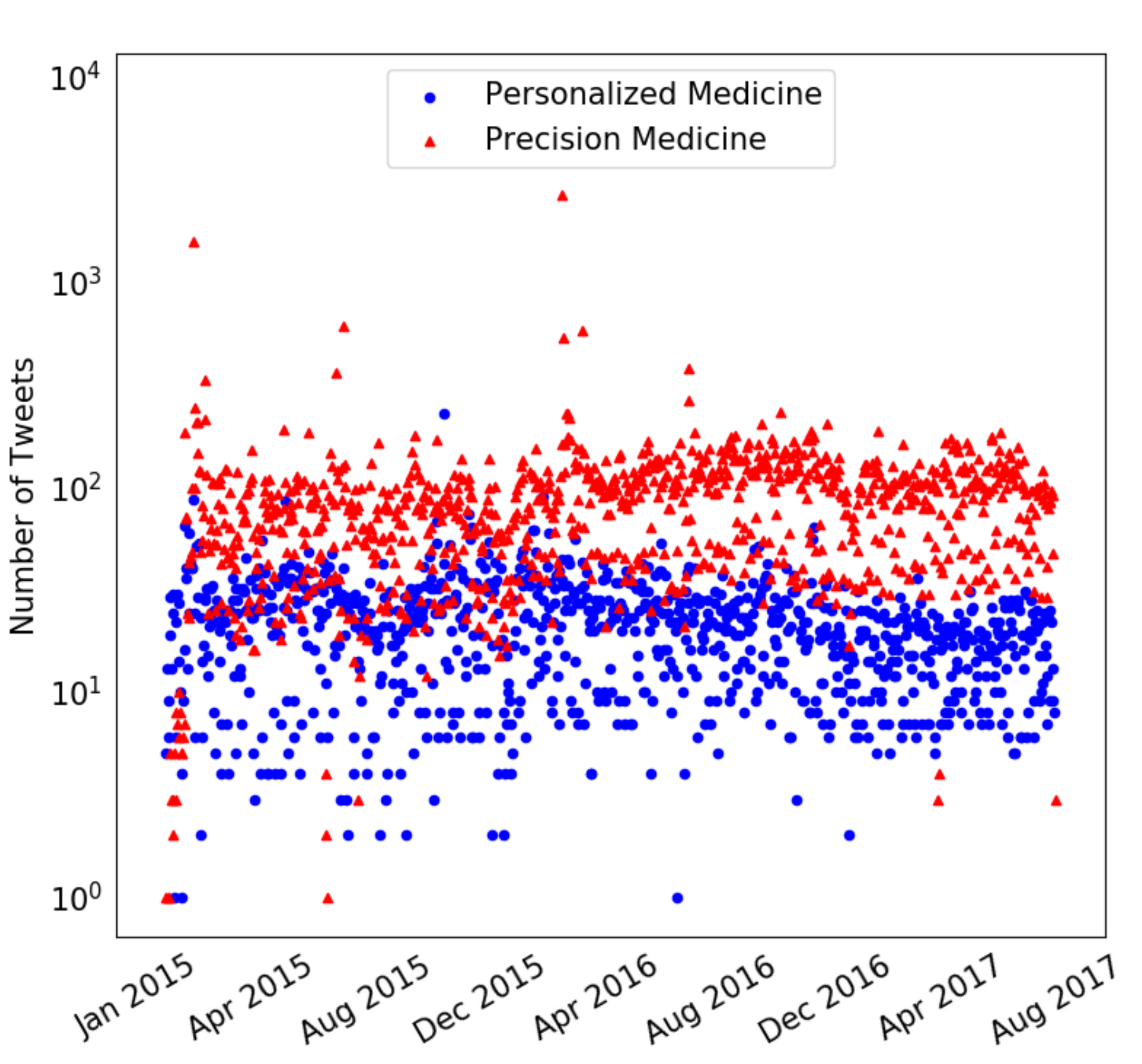}\label{fig:temp-initiatives}}
\caption{Number of tweets per day. Note the log scale in y-axis.}
\label{fig:timeline_analysis}\reduce\reduceB
\end{figure}
}

\descr{Temporal analysis.} %
Finally, we analyze how the volume of tweets changes over time\longVer{. In Figure~\ref{fig:timeline_analysis}, we plot the number of tweets per day in our dataset (between Jan 1, 2015--July 31, 2017)}
for the two most popular companies (23andMe/AncestryDNA) and the two most popular genomics initiatives (Personalized/Precision Medicine).
On average, there are 145 and 30 tweets per day for 23andMe and AncestryDNA keywords, respectively.
\shortVer{(Plots are omitted due to space limitations.)}
While the former is relatively constant, the latter increases steadily in 2017\shortVer{.}\longVer{(Figure~\ref{fig:temp-dtc}).}
This may be the result of AncestryDNA's aggressive promotion strategies (see Sec.~\ref{sec:content-analysis}). %
We also find a number of outliers for 23andMe, mostly around Feb 20 and Oct 19, 2015, and Apr 6, 2017,
which are key dates related to 23andMe's failure to get FDA approval for their health reports in 2015, then obtained in 2017. %
In fact, 20K/132K 23andMe tweets are posted around those dates. 
As for Personalized and Precision Medicine\longVer{ (Figure~\ref{fig:temp-initiatives})}, the volume of tweets stays relatively flat. %
There are outliers for Precision Medicine too, e.g., 2,628 tweets on February 25, 2016, when the White House hosted the Precision Medicine Initiative summit~\cite{whsummit}.

\descr{Discussion.} Overall, our characterization shows that highly engaged users drive the discussion around public genomics initiatives, which is particularly influenced, at least in terms of volumes, by important announcements such as the one made by President Obama.
As for direct-to-consumer (DTC) genetic testing, the conversation is, as expected, dominated by the two most popular companies: 23andMe and AncestryDNA. 
However, it is interesting that the former generates 4 times more tweets even though the latter has more than twice the customers. 
Some of this ``popularity'' seems to be due to 23andMe's controversy around FDA approval. 
We also find a non-negligible use of hashtags, possibly used for promotion and marketing efforts, and that a lot of tweets include URLs to popular domains, indicating that they are used to disseminate news and links to external resources. 
This warrants further exploration, thus, we perform hashtag and URL analysis in the next section.

\begin{table*}[t]
\centering
\setlength{\tabcolsep}{0.5em} %
\footnotesize
\renewcommand{\arraystretch}{0.85}
\resizebox{1\textwidth}{!}{%
\begin{tabular}{lrlrclr}
\toprule
& & \multicolumn{2}{c}{\bf -- Without Official Accounts --} & &  \multicolumn{2}{c}{\bf -- Only Official Accounts --} \\ %
{\bf Keyword} & {\bf WH} & {\bf Top 3 Hashtags} & {\bf KH} & & {\bf Top 3 Hashtags} & {\bf KH} \\
\midrule
23andMe               & 27.09\% & dna (3.58\%), genetics (2.07\%), tech (1.96\%)                    & 12.46\%   & & 23andMestory (6.67\%), genetics (6.35\%), video (5.19\%)                 & 9.74\%  				\\
AncestryDNA           & 75.48\% & sweepstakes (12.38\%), dna (4.90\%), genealogy (4.86\%)           & 25.94\% & & dna (11.74\%), ancestry (5.92\%), familyhistory (5.07\%)                 & 46.88\% 				\\
Counsyl               & 45.24\% & getaheadofcancer (2.64\%), cap (1.93\%), medical (1.94\%)         & 3.08\%  & & acog17 (6.18\%), womenshealthweek (5.15\%), teamcounsyl (5.15\%)\hspace*{-0.1cm}         & 0\%    \\
DNAFit                & 55.30\% & diet (4.19\%), fitness (3.72\%), crossfit (3.54\%)                & 22.91\% & & dna (5.33\%), fitness (3.71\%), generictogenetic (3.48\%)                & 40.37\% 				\\
FamilyTreeDNA         & 29.31\% & dna (14.24\%), genealogy (13.42\%), ancestryhour (3.18\%)         & 10.86\%   & & geneticgenealogy (5.55\%), ftdnasuccess (4.44\%), ftdna (3.33\%)         & 56.66\%  			\\
FitnessGenes          & 72.19\% & startup (5.93\%), london (5.73\%), job (5.59\%)                   & 18.22\% & & fitness (5.85\%), dna (4.32\%), gtsfit (2.79\%)              & 45.29\%  							\\
MapMyGenome           & 54.98\% & shechat (7.94\%), appguesswho (5.32\%), genomepatri (4.22\%)      & 15.80\% & & genomepatri (7.28\%), knowyourself (4.04\%), genetics (2.02\%)           & 0\%     				\\
PathwayGenomics       & 55.85\% & coloncancer (6.91\%), genetictesting (3.29\%), cancer (2.85\%)    & 3.34\%  & & dnaday16 (9.67\%), ashg15 (9.67\%), health (3.22\%)                    & 19.35\%     				\\
Ubiome                & 28.57\% & microbiome (13.23\%), tech (2.14\%), vote (2.07\%)                & 6.61\%  & & microbiome (24.48\%), bacteria (4.76\%), meowcrobiome (2.72\%)           & 6.12\%     			\\
VeritasGenetics       & 57.16\% & brca (3.92\%), genome (3.62\%), genomics (3.32\%)                 & 4.22\%  & & brca (11.82\%), liveintheknow (11.82\%), wholegenome (10.75\%)           & 0\%     				\\ %
Genomics England      & 62.05\% & genomes100k (14.84\%), genomics (7.72\%), raredisease (5.24\%)    & 1.77\%  & & genomes100k (32.45\%), raredisease (19.49\%), genomics (18.71\%)         & 0\%     				\\
Personalized Medicine\hspace*{-0.1cm} & --~    & precisionmedicine (22.74\%), genomics (9.77\%), pmcon (8.37\%)    & --~   & & --~ & --~                                         									\\
Precision Medicine    & --~     & genomics (6.70\%), personalizedmedicine (5.49\%), cancer (4.89\%)\hspace*{-0.3cm} & --~     & & --~ & --~                                       									\\ \bottomrule
\end{tabular}
}
\caption{Top 3 hashtags for each keyword, along with the percentage of tweets with at least a hashtag (WH) as well as that of of ``keyword hashtags'' (KH), e.g., \#23andMe.}
\label{tab:hashtag_analysis}
\vspace*{-0.1cm}
\end{table*}

\section{Content Analysis}\label{sec:content-analysis}
Next, we analyze the content of the tweets related to genetic testing, studying hashtags and URLs included in them and performing a simple sentiment analysis.
We also performed Latent Dirichlet Allocation (LDA)~\cite{blei2003latent} and Term Frequency-Inverse Document Frequency (TF-IDF)~\cite{salton1986introduction} analysis, however, we omit the results since they do not yield any actionable findings.

\begin{table*}[t]
\centering
\setlength{\tabcolsep}{0.5em} %
\footnotesize
\ifshort
  \renewcommand{\arraystretch}{0.85}
\fi
\resizebox{1\textwidth}{!}{%
\begin{tabular}{lll}
\toprule
{\bf Keyword} & {\bf Without Official Accounts} & {\bf Only Official Accounts} \\
\midrule
23andMe               & 23andMe.com (7.33\%), techcrunch.com (3.09\%), fb.me (2.48\%)                   & 23me.co (50.88\%), 23andMe.com (21.13\%), instagram.com (5.40\%)            	\\
AncestryDNA           & journeythroughhistorysweeps.com (15.18\%), ancestry.com (13.94\%),          	& ancstry.me (74.11\%), youtube.com (3.27\%), ancestry.com.au (2.88\%)       	\\
            		  & ancstry.me (6.67\%) 																																			\\
Counsyl               & techcrunch.com (8.42\%), businesswire.com (5.30\%), bioportfolio.com (4.46\%)   & businesswire.com (14.78\%), counsyl.com (13.91\%), medium.com (5.21\%)       	\\
DNAFit                & fb.me (15.81\%), instagram.com (14.65\%), dnafit.com (2.99\%)                   & fb.me (11.74\%), dnafit.com (10.52\%), dnafit.gr (2.83\%)                     \\
FamilyTreeDNA         & familytreedna.com (11.31\%), myfamilydnatest.com (4.28\%), fb.me (4.17\%)       & familytreedna.com (76.56\%), abcn.ws (3.12\%), instagram.com (1.56\%)         \\
FitnessGenes          & instagram.com (14.77\%), fitnessgenes.com (8.48\%), workinstartups.com (6.29\%) & fitnessgenes.com (31.11\%), instagram.com (4.44\%), pinterest.com (4.44\%)    \\
MapMyGenome           & yourstory.com (11.84\%), owler.us (11.44\%), mapmygenome.in (9.18\%)            & mapmygenome.in (42.12\%), youtu.be (14.35\%), indiatimes.com (3.70\%)         \\
PathwayGenomics       & paper.li (11.96\%), atjo.es (10.82\%), pathway.com (3.31\%)                     & pathway.com (23.07\%), nxtbook.com (3.84\%), drhoffman.com (3.84\%)           \\
Ubiome                & techcrunch.com (9.30\%), bioportfolio.com (4.83\%), ubiomeblog.com (4.21\%)     & ubiomeblog.com (34.32\%), igg.me (26.07\%), ubiome.com (6.60\%)               \\
VeritasGenetics       & veritasgenetics.com (10.97\%), technologyreview.com (5.01\%), buff.ly (2.30\%)  & veritasgenetics.com (75.67\%), biospace.com (1.35\%), statnews.com (1.35\%)   \\ %
Genomics England      & genomicsengland.co.uk (33.85\%), youtube.com (1.98\%), buff.ly (1.64\%)         & genomicsengland.co.uk (98.03\%), peoplehr.net (0.58\%),						\\
            		  & 																				& campaign-archive1.com (0.21\%)												\\
Personalized Medicine & instagram.com (8.78\%), myriad.com (2.54\%), buff.ly (2.32\%)                   & --~                                                       					\\
Precision Medicine    & buff.ly (2.92\%), instagram.com (2.27\%), nih.gov (1.87\%)                      & --~                                                     						\\ \midrule
Baseline        	  & instagram.com (4.18\%), fb.me (3.44\%), youtu.be (2.72\%)                     	& --~                                                   						\\ \bottomrule
\end{tabular}
}
\vspace{0.1cm}
\caption{The top 3 domains per keyword, without official accounts and only considering the official accounts.}
\label{tab:top_domains}\reduce\reduceA
\end{table*}

\reduceB
\subsection{Hashtag Analysis}\label{sec:keyword-dataset-hashtag}
In Table~\ref{tab:hashtag_analysis}, we report the top three hashtags for every keyword, while differentiating between tweets made by regular users and those by official accounts. 
We also quantify the percentage of tweets with at least one hashtag (WH) and \longVer{that of tweets }including the keyword as a hashtag (KH), e.g., \#23andMe.

We find a few unexpected hashtags among the DTC tweets, e.g., \#sweepstakes (AncestryDNA), \#startup (Fitnessgenes), \#vote (Ubiome), \#shechat and \#appguesswho (MapMyGenome).
AncestryDNA's top hashtag, \#sweepstakes (12\%), is related to a marketing campaign promoting a TV series, 
``America: Promised Land.'' 
There are 3.5K tweets, from distinct users, with the very same content (most likely due to a ``share'' button): \enquote{I believe I've discovered my @ancestry! Discover yours for the chance to win an AncestryDNA Kit. \#sweepstakes \url{journeythroughhistorysweeps.com}.}
We also find hashtags like \#feistyfrugal and \#holidaygiftguide in the AncestryDNA top 10 hashtags, which confirms how AncestryDNA uses Twitter for relatively aggressive marketing campaigns.
Moreover, in the Fitnessgenes tweets, we find hashtags like \#startup, \#london, and \#job due to a number of tweets advertising jobs for Fitnessgenes, while \#shechat appears in tweets linking to an article related to women in business about MapMyGenome's founder.

By contrast, top hashtags for official accounts' tweets are closer to their main expertise/business. 
Similarly, those for genomics initiatives are pretty much always related to genetic testing, and this is actually consistent besides top 3.
\longVer{(The top 10 hashtags include, e.g., \#digitalhealth, \#genetics, and \#lifestylemedicine).}
Finally, the percentage of tweets with the keyword appearing as a hashtag (KH), range from 12\% for 23andMe to 25\% for AncestryDNA even when excluding official accounts, which might be the by-product of promotion campaigns.
When looking at tweets by official accounts KH values go up for some companies, e.g., AncestryDNA heavily promotes their brand using hashtags (46\% KH). 

\subsection{URL Analysis}\label{sec:content-url}

Next, we analyze the URLs contained in the tweets of our dataset.
Recall that the ratio of tweets containing URLs, as well as the percentage of those in the Alexa top 1M domains, are reported in Table~\ref{tab:keyword_dataset}.
Once again, we distinguish between tweets from the official accounts and report the top 3 (top-level) domains per keyword in Table~\ref{tab:top_domains}. 
Since there are several URL shorteners in our dataset (e.g., \url{bit.ly}), %
so we first extract the top 10 domains for each keyword and identify those {\em only} providing URL shortening services, then, we ``unshorten'' the URLs and use them in our analysis instead.

Among the top URLs shared by the official accounts, we find, unsurprisingly, their websites, as well as others leading to other domains owned by them, e.g., \url{23me.co}, \url{ancestry.com.au}, and \url{ancstry.me}. 
A few companies also promote news articles about them or related topics, e.g., top domains for Counsyl and MapMyGenome include \url{businesswire.com} and \url{indiatimes.com},
while DNAfit seems more focused on social media with its top domain being Facebook. 
As discussed previously, the domain \url{journeythroughhistorysweeps.com} appears frequently in AncestryDNA tweets. %
Then, note that \url{techcrunch.com}, a blog about technology, appears several times, as it often covers news and stories about genetic testing. 
We also highlight the presence of \url{owler.us}, an analytics/marketing provider sometimes labeled as potentially harmful by Twitter, as one of the top domains for MapMyGenome.

Finally, for genomics initiatives, we notice \url{buff.ly}, a social media manager, suggesting that interested users appear to be extensively scheduling posts, thus potentially being more tech-savvy.
We also find \url{myriad.com}, the domain of Myriad Genetics, which discovered the BRCA1 gene and tried to patent it~\cite{caulfield2007myriad}, as their account is quite active in posting tweets with Precision/Personalized Medicine keywords.\reduceA

\subsection{Sentiment Analysis}\label{sec:content-sentiment}

\longVer{
\begin{figure}[t]
\vspace{0.3cm}
\center
\includegraphics[width=0.34\textwidth]{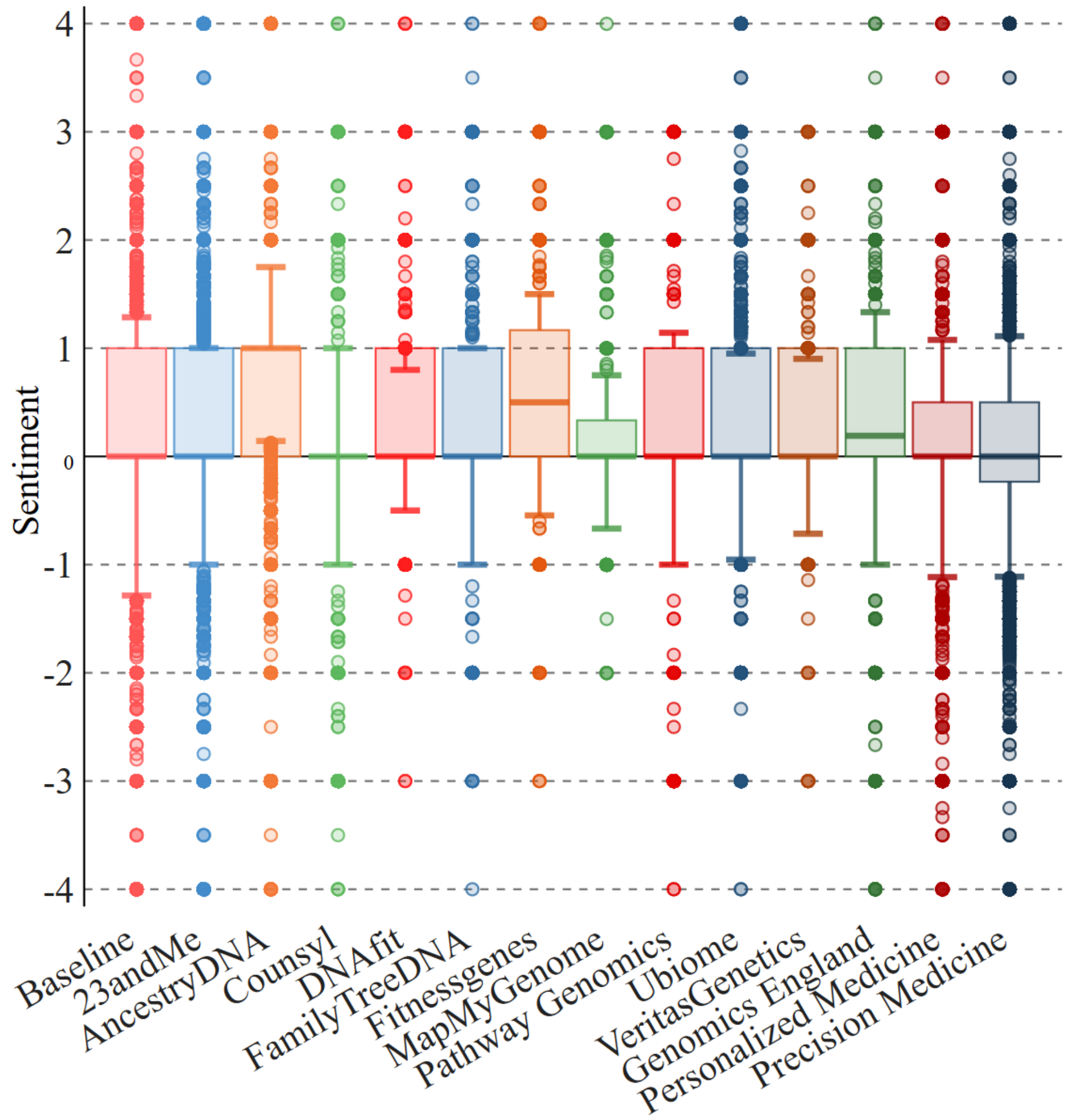}
\caption{Sentiment scores of the keyword dataset.}
\label{fig:sentiment_analysis_keyword_boxes}\reduce
\end{figure}}

\shortVer{
\begin{figure*}[t]
\center
\subfigure[followers]{\includegraphics[width=0.213\textwidth]{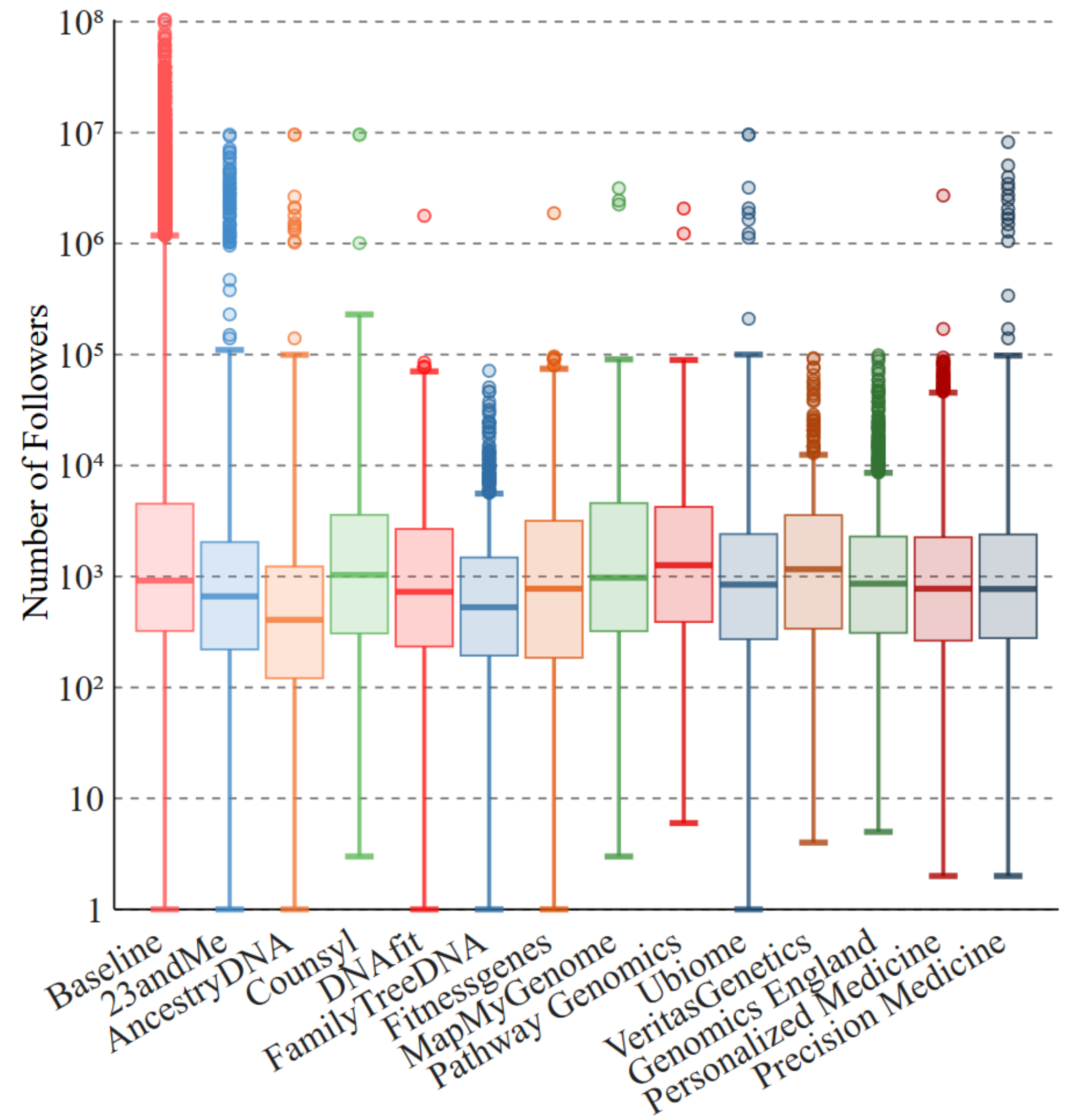}\label{fig:number_of_followers}}
\subfigure[following]{\includegraphics[width=0.213\textwidth]{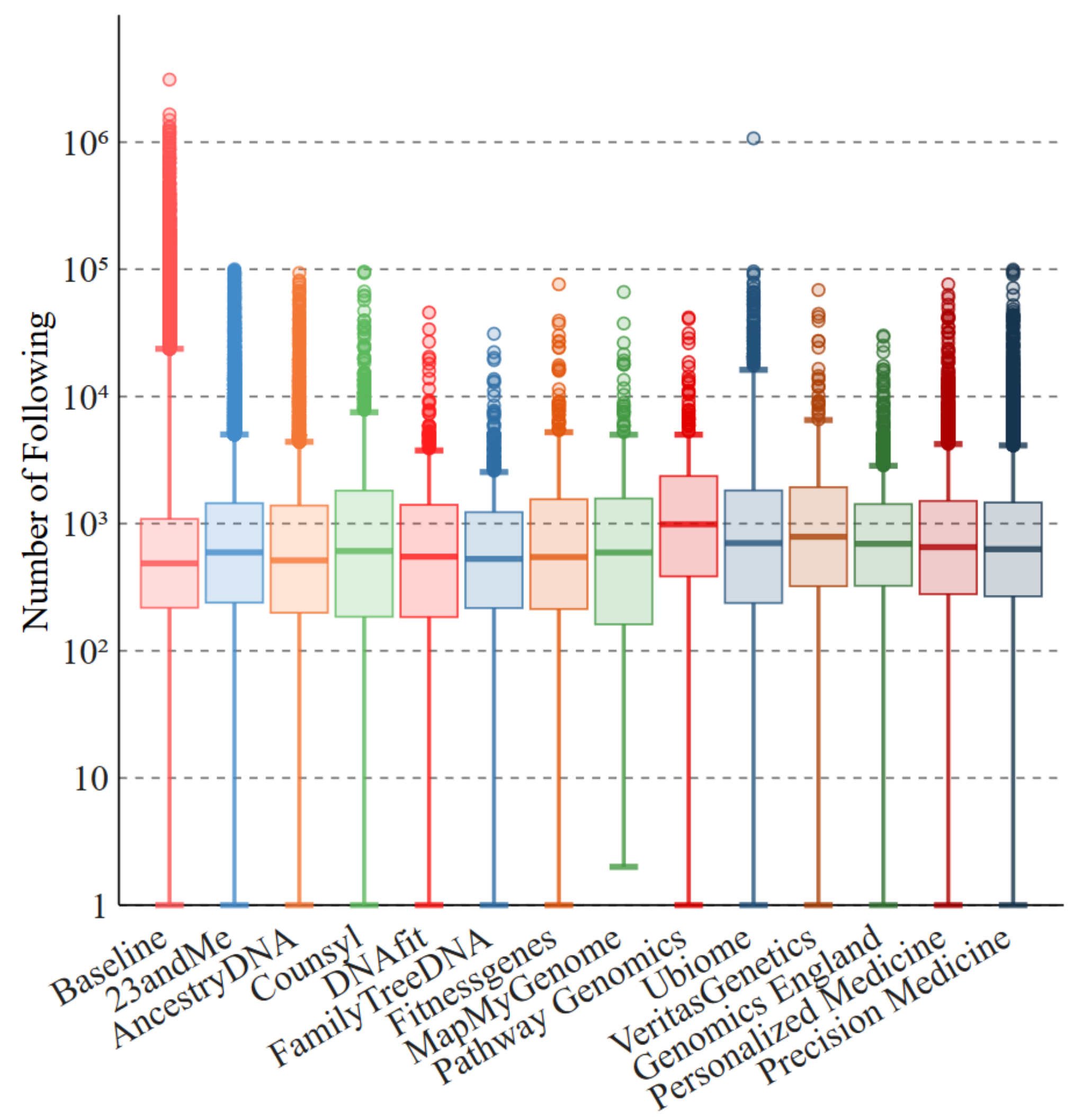}\label{fig:number_of_following}}
\subfigure[likes]{\includegraphics[width=0.213\textwidth]{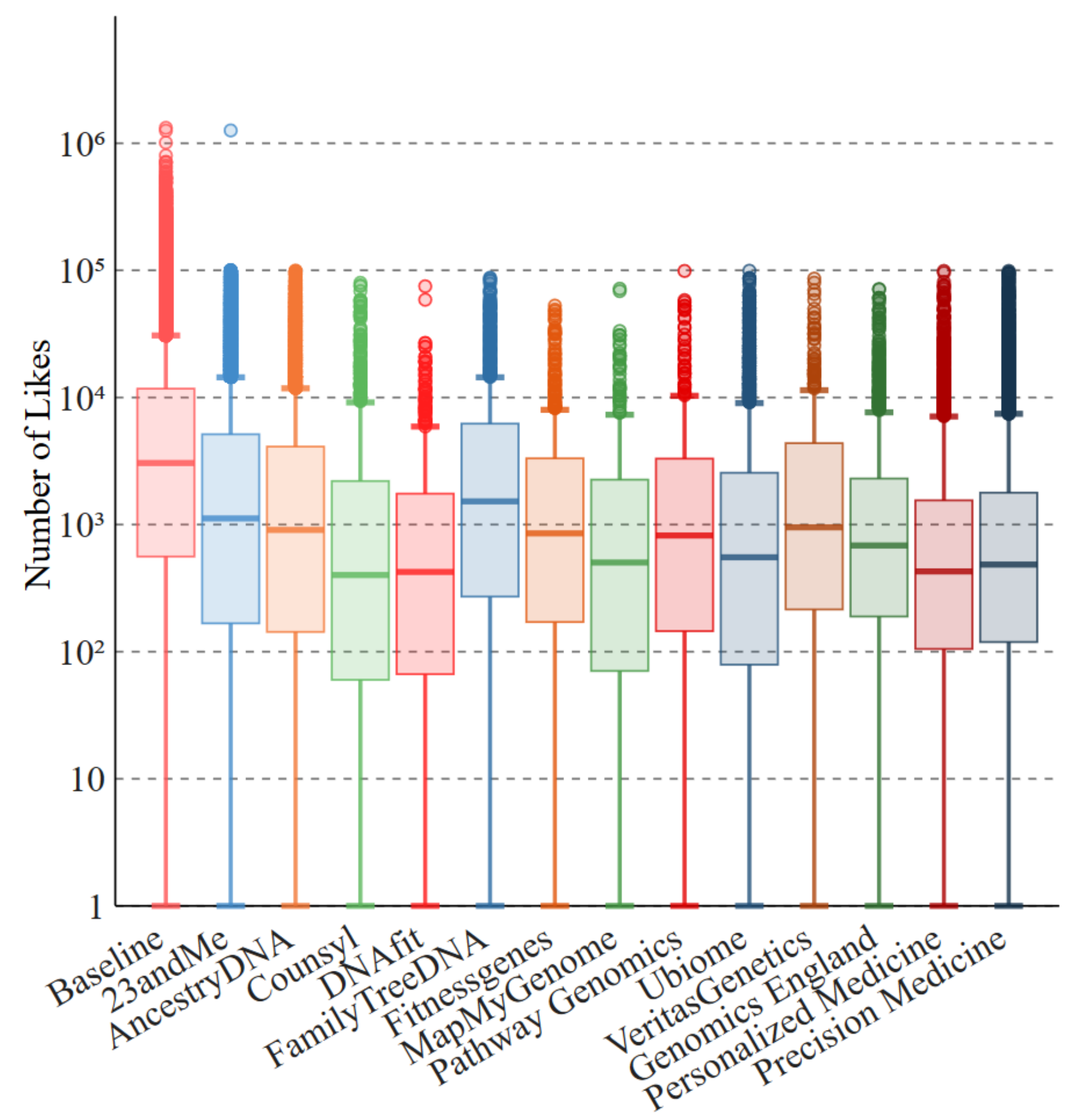}\label{fig:number_of_likes}}
\subfigure[tweets]{\includegraphics[width=0.213\textwidth]{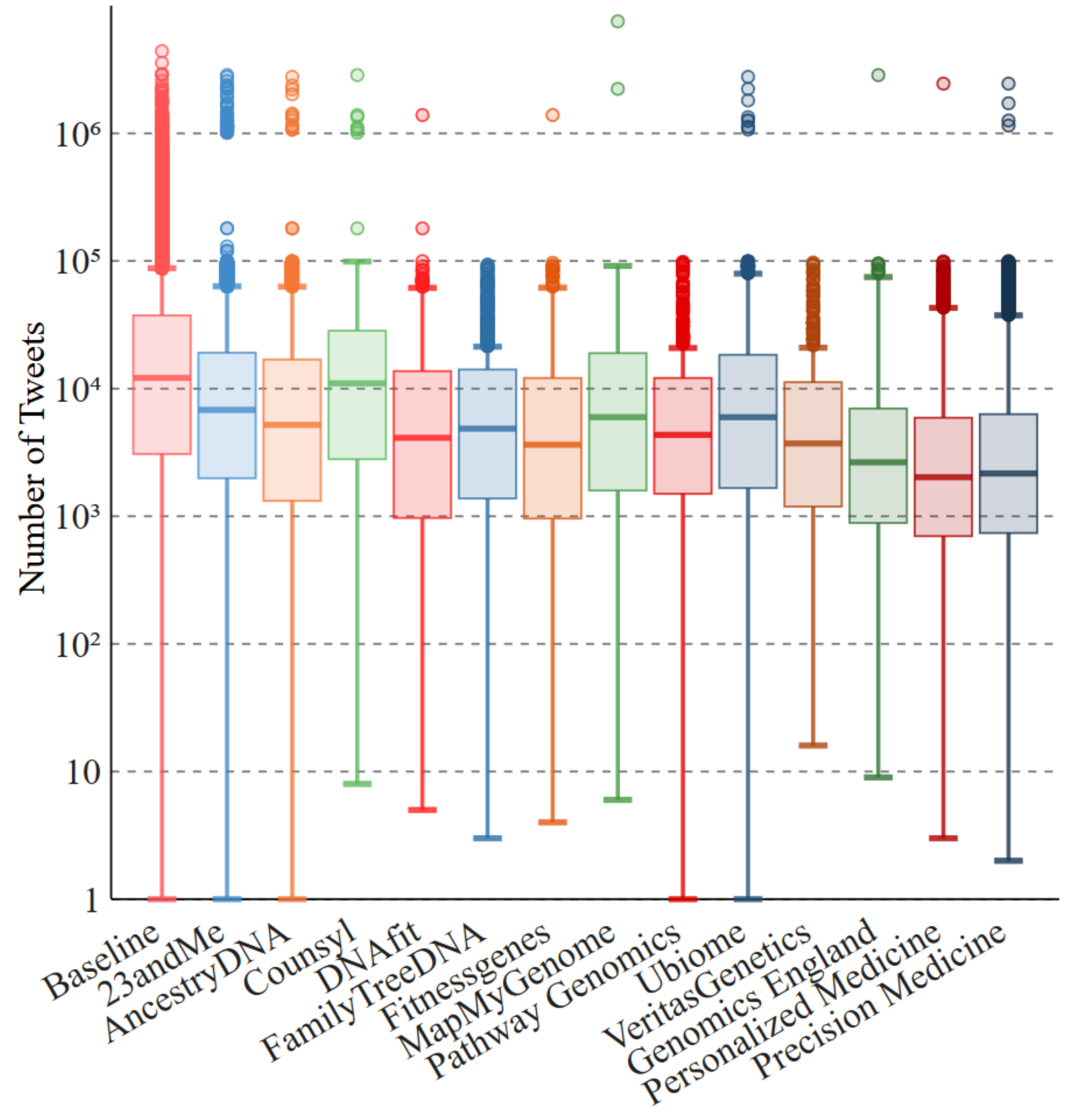}\label{fig:number_of_tweets}}
\caption{Boxplots with statistics per user profile (note the log-scale in y-axis).}
\label{fig:popularity}\reduce\reduce
\end{figure*}}

\longVer{
\begin{figure*}[t]
\center
\subfigure[followers]{\includegraphics[width=0.24\textwidth]{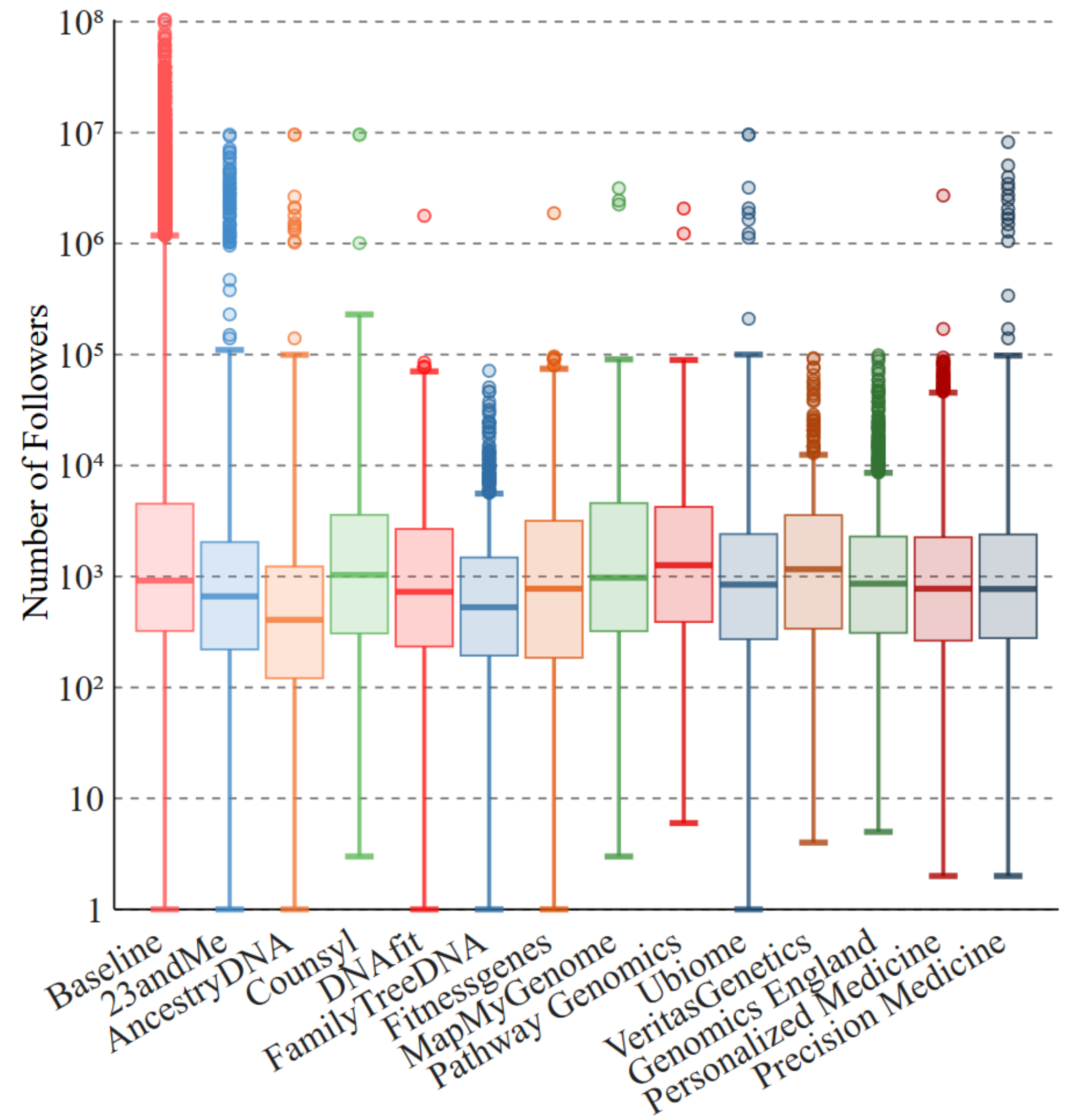}\label{fig:number_of_followers}}
\subfigure[following]{\includegraphics[width=0.24\textwidth]{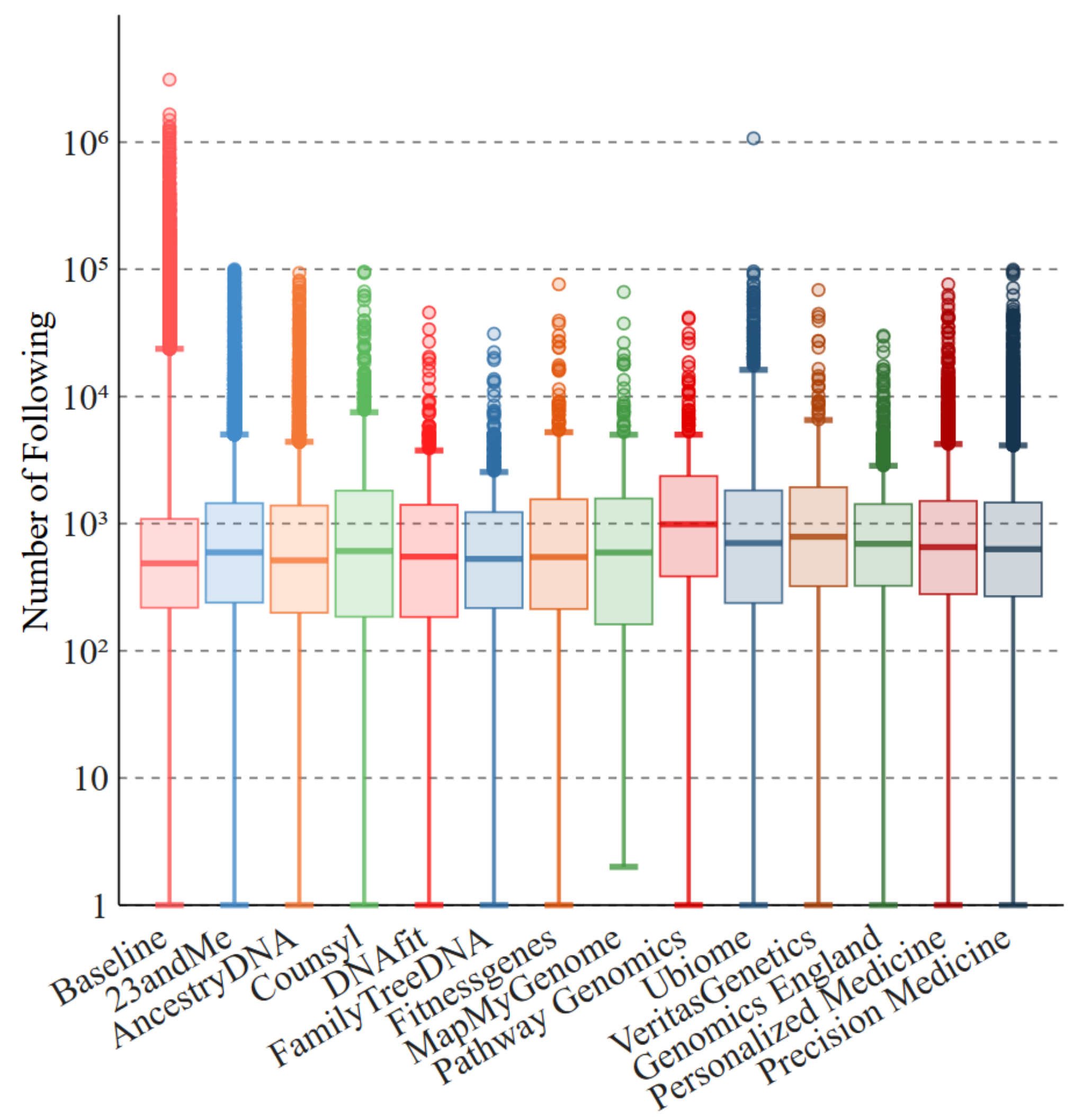}\label{fig:number_of_following}}
\subfigure[likes]{\includegraphics[width=0.24\textwidth]{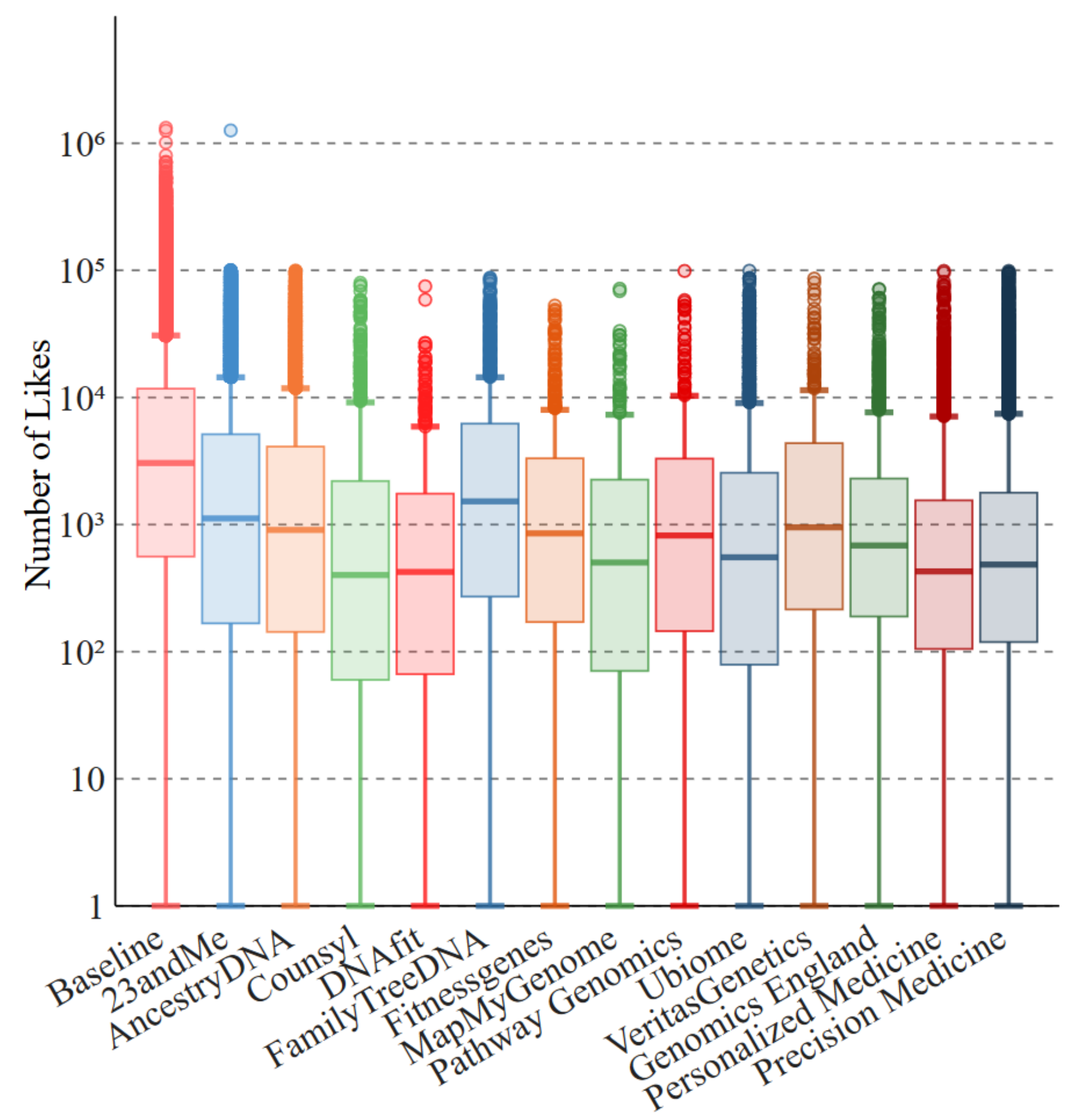}\label{fig:number_of_likes}}
\subfigure[tweets]{\includegraphics[width=0.24\textwidth]{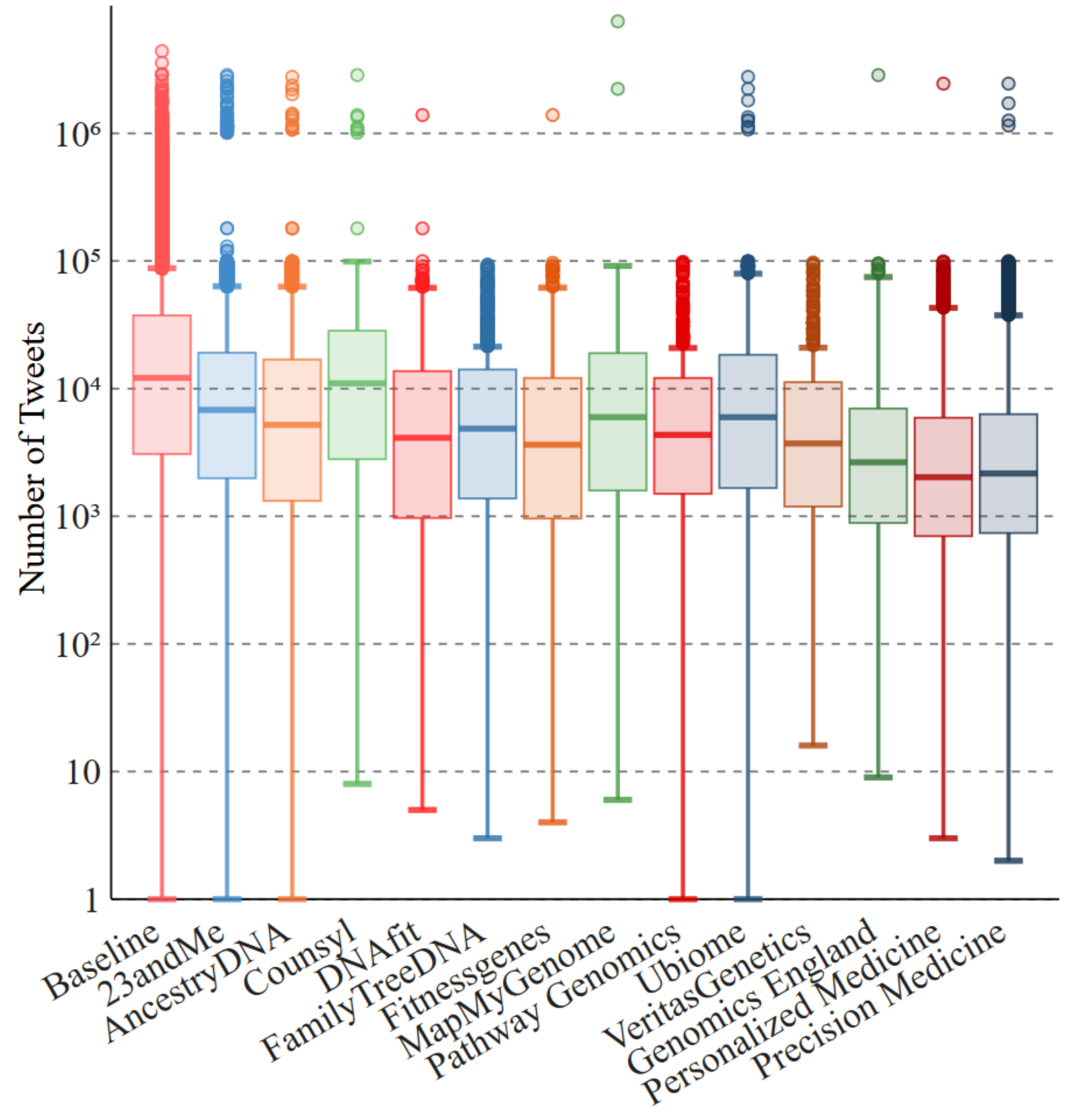}\label{fig:number_of_tweets}}
\caption{Boxplots with statistics per user profile (note the log-scale in y-axis).}
\label{fig:popularity}
\vspace{-0.2cm}
\end{figure*}
}

We perform sentiment analysis using SentiStrength~\cite{thelwall2010sentiment}\longVer{, which is designed to work on short texts}.
The tool outputs two scores, one positive, in $[1,5]$, and one negative, in $[-1,-5]$.
We calculate the sum value of the positive+negative scores for every tweet, then, collect {\em all} tweets with that keyword from the {\em same} user, and output the mean sentiment score.

\longVer{In Figure~\ref{fig:sentiment_analysis_keyword_boxes}, we report the distribution of sentiment across the different keywords.}
The vast majority of tweets have neutral sentiment, ranging from 0 to 1 scores.
We run pair-wise two-sample Kolmogorov-Smirnov tests on the distributions, and in most cases reject the null hypothesis that they come from a common distribution at $\alpha=0.05$.
However, we are \emph{unable} to reject the null hypothesis when comparing the baseline dataset to the PathwayGenomics dataset ($p = 0.77$) and when comparing DNAfit to Ubiome ($p = 0.34$). %
In general, the genomics initiatives, and in particular Personalized Medicine and Precision Medicine, have many outliers compared to most DTC genetic companies, suggesting more users who reveal strong feelings for or against these concepts. 
Genomics England, however, has a median above zero, indicating generally positive sentiment. 
Tweets about Counsyl are very neutral, while Ubiome tweets seem to be the most positive.

\subsection{Discussion}
Our content analysis yields a few interesting findings. 
A large part of the genetic testing discourse appears to be generated from news and technology websites, and from tech-savvy users who rely on services to schedule social media posts. 
Also, sentiment around DTC companies is overall neutral, but positive for the genomics initiatives, however, tweets about DTC companies include a lot of strongly opinionated users (both positive and negative)\shortVer{.}\longVer{; we further explore tweets with high negative score in Sec.~\ref{sec:use-case}.}
Finally, tweets related to genetic testing not only contain a significantly higher number of hashtags than a random baseline, but they are also used for promotion.
In general, we find several social media marketing strategies at play, with some companies employing traditional giveaways, others promoting mainly third-party articles about the company, and others focusing their efforts across multiple social media platforms.
For instance, AncestryDNA is quite active in this context, with one particular hashtag (\#sweepstakes) found in 1 out of 8 AncestryDNA tweets. 
This has a significant impact on how ``regular'' users engage in tweeting about genetic testing, which we further analyze next.

\section{User Analysis}\label{sec:user-analysis}

In this section, we shed light on the accounts tweeting about genetic testing.
After a general characterization of the profiles, we look for the presence of social bots~\cite{varol2017online}. 
Then, we select a random sample of users tweeting about the two most popular DTC companies and analyze their latest 1,000 tweets to understand their interests.

\subsection{User Profiles}\label{sec:user-statistics}
We start by analyzing the profiles tweeting about genetic testing: in Figure~\ref{fig:popularity}, we plot the distribution of the number of their followers, following, likes, and tweets.

\descr{Followers.} Accounts tweeting about genomics initiatives have a median number of followers similar to baseline, while for the DTC companies the median is always lower, except for Counsyl, MapMyGenome, PathwayGenomics, and VeritasGenetics (see Figure~\ref{fig:number_of_followers}).
Also considering that, for these four companies, there is a relatively low number of unique users (see Table~\ref{tab:keyword_dataset}), we believe accounts tweeting about them are fewer but more ``popular.''
There are fewer outliers than the baseline, which is not surprising since we do not expect many mainstream accounts to tweet about genetic testing. 
Some outliers appear for 23andMe and AncestryDNA, which, upon manual examination, turn out to be Twitter accounts of newspapers or known technology websites, reflecting how the two most popular companies also get more press coverage.

\descr{Following.} The median number of `following' (i.e., the accounts followed by the users in our dataset) is usually higher than baseline for DTC companies but similar for genomics initiatives (Figure~\ref{fig:number_of_following}).
This suggests that users interested in DTC genetic testing might want to get more information off Twitter and/or from more accounts.

\descr{Likes.} We then measure the number of tweets each profile has liked (Figure~\ref{fig:number_of_likes}). 
This measure, along with the number of tweets, depicts, to a certain extent degree, a level of engagement.
We find that, for all keywords, profiles like fewer tweets than baseline users.
There is one interesting outlier for 23andMe (@littlebytesnews), who liked more than 1M tweets; 
this is likely to be a bot, as also confirmed by Botometer~\cite{varol2017online}.
Also, FamilyTreeDNA appears to have users liking more tweets than others. %
However, these accounts appear not to be bots, as we discuss later.

\descr{Tweets.} We also quantify the number of tweets each account posts (Figure~\ref{fig:number_of_tweets}).
As with the number of likes, users in our datasets are less ``active'' than baseline users.
There are interesting outliers above 1M tweets, which are due to social bots. 
We also find more tweets from Counsyl's users, seemingly mostly due to a large number of profiles describing themselves as ``promoters'' of science/digital life, technology enthusiasts, and/or influencers. 
Finally, users tweeting about genomics initiatives appear to be even less active, with a lower median value of tweets than the rest.
Also considering that these users tweet more about the same keyword (as discussed in Sec.~\ref{sec:content-general}) but follow more accounts, we believe that they are \longVer{somewhat }more {\em passive} than the average Twitter user, \longVer{possibly }using Twitter to get information but actively engaging less than others.

\begin{figure}[t]
\centering
\shortVer{\includegraphics[width=0.63\columnwidth]{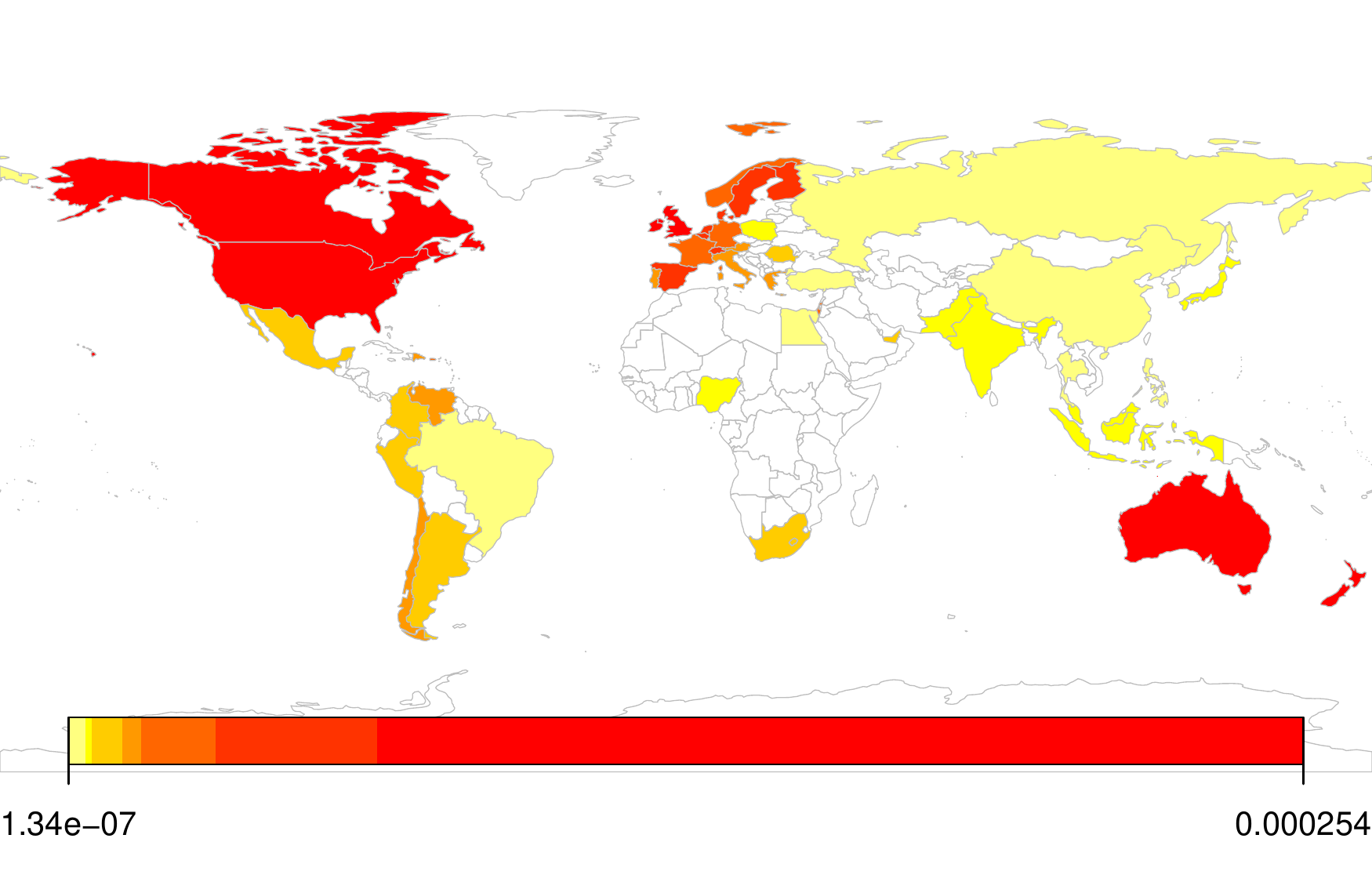}}
\longVer{\includegraphics[width=0.775\columnwidth]{plots/twitter-map.pdf}}
\caption{Geolocation of Twitter profiles, normalized by Internet using population per country.}
\label{fig:geolocation2}\reduceA\reduceA
\reduce
\end{figure}

\descr{Geographic Distribution.} We then estimate the geographic distribution of the users via the location field in their profile.
This is self-reported, and users use it in different ways, adding their city (e.g., Miami), state (e.g., Florida), and/or country (e.g., USA).
In some cases, entries might be empty (7.5\% of the tweets in our dataset), ambiguous (e.g., Paris, France vs Paris, Texas), or fictitious (e.g., ``Hell''). 
Nevertheless, as done in previous work~\cite{marcus2011twitinfo}, we use this field to estimate where most of the tweets are coming from.
We use the Google Maps Geolocation API, which allows to derive the country from a text containing a location.\longVer{\footnote{\url{https://developers.google.com/maps/documentation/geolocation}}}
The API returns an error for 6.6\% of the profiles, mostly due to fictitious locations.

We find that the top 5 countries in our dataset are mostly English-speaking ones: 69.1\% of all profiles with a valid location are from the US, followed by the UK (8.6\%), Canada (4.5\%), India (2.1\%), and  Australia (1.4\%).
We then {\em normalize} using Internet-using population estimates~\cite{internetStats}, and plot the resulting heatmap, with the top 50 countries, in Figure~\ref{fig:geolocation2}.
The maximum value is obtained by the US (i.e., 0.000254 users per Internet user), with 72.8K unique users, out of an estimated Internet population of 286M, posting tweets in our dataset.
This suggest that US users dominate the conversation\shortVer{.}\longVer{ on genetic testing on Twitter.}

We also perform a geolocation analysis broken down to specific keywords.
Unsurprisingly, the top country of origin for Genomics England is the UK, as it is for DNAfit, which is based in London.
Similarly, the top country for India-based company MapMyGenome tweets is India.
Overall, we find that tweet numbers are in line with the countries where the DTC companies are based or operate -- e.g., 23andMe health reports are available in US, Canada, and UK, while AncestryDNA also operates in Australia -- as well as where the genomics initiatives are taking place.

\subsection{Social Bot Analysis}\label{sec:bot}

Next, we investigate the presence of social bots in our datasets, using the Botometer (\url{botometer.iuni.iu.edu}), a tool that, given a Twitter handle, returns the probability of it being a ``social bot,'' i.e., an account controlled by software, algorithmically generating content and establishing interactions~\cite{varol2017online}. 

\longVer{
\begin{figure}
\vspace{0.4cm}
\center
\includegraphics[width=0.34\textwidth,height=5.0cm]{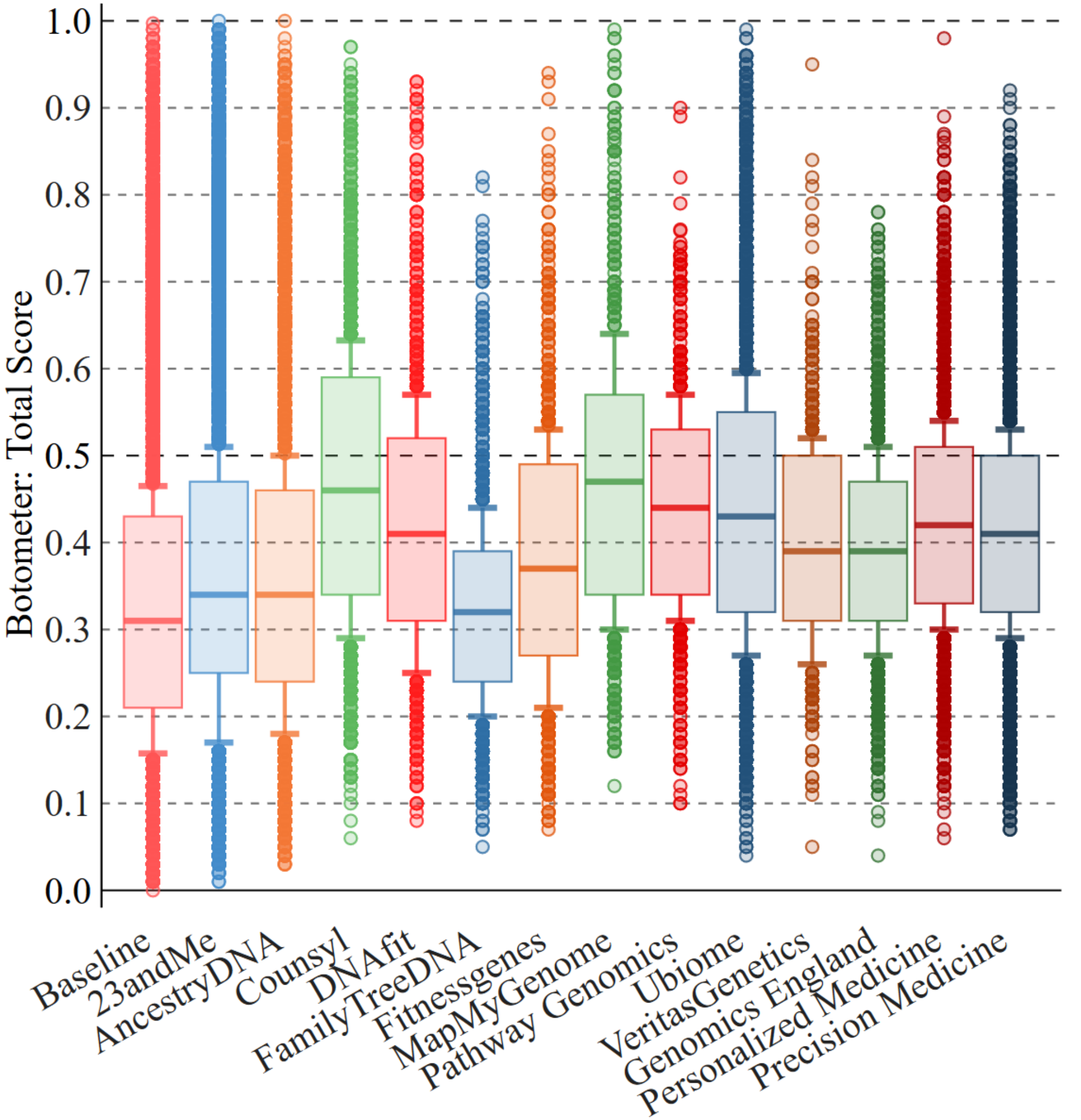}\label{fig:botometer_total_score}
\caption{Botometer scores for the keyword dataset.} %
\label{fig:botometer}\reduce
\end{figure}
}

\longVer{In Figure~\ref{fig:botometer}, we plot the distribution of Botometer scores for all keywords. We}
\shortVer{Although actual plots are included in the extended version of the paper~\cite{full}, here we}
compare the distributions using pairwise 2 sample KS tests, and reject the null hypothesis at $\alpha = 0.05$ for all datasets \emph{except} Counsyl and MapMyGenome ($p = 0.29$), DNAfit and VeritasGenetics ($p = 0.17$) and PrecisionMedicine and VeritasGenetics ($p = 0.10$).
We also find that all median scores are higher than the baseline (between 0.35 and 0.5 vs 0.3).
This is not entirely surprising since we expect many blogs, magazines, and news services covering genetic testing, and these are likely to get higher scores than individuals since they likely automate their activities. 
However, about 80\% of the accounts in our dataset have scores lower than 0.5 and 90\% lower than 0.6 (i.e., it is unlikely they are bots).
We also find the two most popular keywords, 23andMe and AncestryDNA, as well as FamilyTreeDNA, somewhat stand out: accounts tweeting about them get the lowest Botometer scores.
Although for FamilyTreeDNA this might be an artifact of the relatively low number of tweets (2K users),
the scores suggest there might be more interaction/engagement from ``real'' individuals and/or fewer tweets by automated accounts about 23andMe and AncestryDNA.

We then look at accounts with Botometer scores \emph{above} 0.7, finding that, for most DTC keywords, they account for 3--5\% of the users; not too far from the baseline (2\%) and the genomics initiatives (1.5--2\%). 
Counsyl and MapMyGenome have more than 10\% of users with scores above 0.7. 
We also quantify {\em how many} tweets are posted by (likely) social bots: almost 15\% of all PathwayGenomics tweets come from users with score 0.7 or above (4.5\% of all users), while for all other keywords social bots are not responsible for a substantially high number of tweets in our datasets.

\subsection{Last 1K Tweets of a Sample of Users}

We then focus on the users tweeting about the two most popular companies -- i.e., 23andMe and AncestryDNA -- and study their last 1K tweets aiming to understand the characteristics of the accounts who show interest in genetic testing. %
\longVer{We only do so for 23andMe and AncestryDNA as these companies have the highest numbers of tweets and users, and thus, are more likely to lead to a representative and interesting sample.}

\descr{Data Crawl.} We select a random 20\% sample of the users who have posted at least one tweet with keywords 23andMe/AncestryDNA (resp., 12.2K/64K and 3.3K/16.9K users) and crawl their latest 1K tweets if their account is still active.\footnote{We find 575 and 61 inactive accounts, resp., for 23andMe and AncestryDNA.}
This yields a dataset of 12M tweets, outlined in Table~\ref{tab:user_dataset}.
For comparison, we also get the last 1K tweets of a random sample of 5K users from the keyword dataset's baseline users.
Note that statistics in Table~\ref{tab:user_dataset} refer to the latest 1K tweets of the user sample, while those in Table~\ref{tab:keyword_dataset} to tweets with a given keyword.

The numbers of retweets and likes per tweet are, once again, lower than the baseline.
However, users tweeting about AncestryDNA receive, for their last 1K tweets, one order of magnitude more likes than those tweeting about 23andMe.
Moreover, we observe relatively high percentages of tweets with hashtags (63\%) and URLs (around 80\%).
How far back in time the 1,000th tweet appears varies across users, depending on how often they tweet.
We measure the time between the most recent and the 1,000th tweet, and find that baseline users are more ``active'' than the users who have tweeted about 23andMe and AncestryDNA, in line with what discussed previously. %
In particular, AncestryDNA users appear to post less: for half of them, it takes at least 359 days to tweet 1K tweets compared to 260 for the baseline and 287 for 23andMe. 

\begin{table}[t]
\centering
\footnotesize
\ifshort
  \renewcommand{\arraystretch}{0.85}
\fi
\setlength{\tabcolsep}{0.3em} %
\resizebox{\columnwidth}{!}{%
\begin{tabular}{lrrrrrrr}
\toprule
\textbf{}       	& \textbf{Tweets} & \textbf{Users} & \textbf{RTs} & \textbf{Likes} & \textbf{Hashtags} & \textbf{URLs} & \textbf{Top 1M}  \\ \midrule
23andMe         	& 9,534,302       & 12,227         & 9,077,066    & 3,501,053      & 24.40\%           & 63.62\%       & 81.43\%          \\
AncestryDNA     	& 2,466,443       & 3,320          & 1,399,804    & 22,001,065     & 34.21\%           & 63.64\%       & 78.86\%          \\ %
{\em Total}         & 12,000,745      & 15,547         & 10,476,870   & 25,502,118     & 26.41\%           & 63.62\%       & 80.89\%          \\ %
{\em Baseline}      & 4,208,967       & 5,035          & 139,551,104  & 342,052,546    & 17.47\%           & 41.24\%       & 88.41\%      	  \\ \bottomrule
\end{tabular}
}
\caption{Summary of the users' tweets dataset, with last 1K tweets of a 20\% sample of 23andMe and AncestryDNA users.}
\label{tab:user_dataset}
\shortVer{\reduce}
\end{table}

\begin{table}[t]
\centering
\footnotesize
\ifshort
  \renewcommand{\arraystretch}{0.85}
\fi
\setlength{\tabcolsep}{0.25em}
\begin{tabular}{lll}
\toprule
\multicolumn{1}{l}{\bf 23andMe}         & \multicolumn{1}{l}{\bf AncestryDNA}   &  {\bf Baseline}             	\\ \midrule
tech (1.07\%)                         	& giveaway (3.31\%)           			& gameinsight (0.55\%)          \\
news (1.06\%)                         	& sweepstakes (2.01\%)          		& trecru (0.34\%)           	\\
health (0.58\%)                       	& win (2.01\%)              			& btsbbmas (0.33\%)           	\\
business (0.48\%)                       & genealogy (1.01\%)            		& nowplaying (0.30\%)         	\\
healthcare (0.43\%)                     & tech (0.63\%)             			& android (0.28\%)            	\\
digitalhealth (0.40\%)                  & ad (0.51\%)               			& androidgames (0.27\%)         \\
startup (0.39\%)                    	& entry (0.51\%)              			& ipad (0.26\%)             	\\
socialmedia (0.34\%)                    & promotion (0.48\%)            		& trump (0.24\%)            	\\
viral (0.34\%)                    		& perduecrew (0.47\%)           		& music (0.21\%)            	\\
technology (0.34\%)                     & contest (0.44\%)            			& ipadgames (0.20\%)          	\\ \bottomrule
\end{tabular}
\caption{The top 10 hashtags of the users' tweets dataset.}
\label{tab:hashtag_analysis_user}
\shortVer{\reduce\reduce}
\end{table}

\descr{Hashtag analysis.}
We conduct a hashtag analysis on tweets in Table~\ref{tab:user_dataset}.
In Table~\ref{tab:hashtag_analysis_user}, we report the top 10 hashtags of the users' last 1K tweets.
For 23andMe, we find several hashtags related to health in the top 10; also considering that the top 30 include \#pharma, \#cancer, and \#biotech, it is likely that users who have shown interest in 23andMe are also very much interested in (digital) health, which is one of the primary aspects of 23andMe's business.
This happens to a lesser extent for AncestryDNA results: while top hashtags include \#genealogy (4th), they also include \#giveaway, \#sweepstakes, \#win, \#ad, \#promotion, \#perduecrew, and \#contest, 
suggesting that these users are rather interested in promotional products. 
This is line with our earlier observation that AncestryDNA extensively uses advertising and marketing campaigns on Twitter. 

\begin{table}[t]
\centering
\footnotesize
\ifshort
  \renewcommand{\arraystretch}{0.85}
\fi
\setlength{\tabcolsep}{0.3em} %
\resizebox{\columnwidth}{!}{%
\begin{tabular}{lll}
\toprule
\multicolumn{1}{l}{\bf 23andMe} &  \multicolumn{1}{l}{\bf AncestryDNA}      &  {\bf Baseline}             	\\ \midrule
fb.me (4.00\%)          		&  instagram.com (6.78\%)             		&  fb.me (5.85\%)           	\\
instagram.com (3.06\%)      	&  fb.me (5.48\%)                 			&  instagram.com (4.42\%)       \\
youtu.be (2.18\%)         		&  techcrunch.com (4.42\%)            		&  youtu.be (2.94\%)          	\\
buff.ly (2.17\%)        		&  youtu.be (4.04\%)              			&  twittascope.com (0.58\%)     \\
techcrunch.com (1.53\%)     	&  wn.nr (1.79\%)                 			&  tmblr.co (0.56\%)          	\\ 
lnkd.in (1.02\%)          		&  woobox.com (1.51\%)              		&  buff.ly (0.54\%)           	\\ 
mashable.com (0.65\%)      	 	&  giveaway.amazon.com (1.17\%)         	&  fllwrs.com (0.40\%)          \\ 
entrepreneur.com (0.63\%)     	&  buff.ly (1.08\%)               			&  gigam.es (0.33\%)          	\\ 
nyti.ms (0.62\%)          		&  swee.ps (0.80\%)               			&  soundcloud.com (0.32\%)      \\ 
reddit.com (0.55\%)       		&  twittascope.com (0.41\%)           		&  vine.co (0.30\%)           	\\ \bottomrule
\end{tabular}
}
\caption{The top 10 domains of the users' tweets dataset.}
\label{tab:top_domains_user}
\shortVer{\reduce}
\end{table}

\descr{URL analysis.} In Table~\ref{tab:top_domains_user}, we report the top 5 domains of the three sets. 
Over the last 1K tweets, users tweeting about 23andMe and AncestryDNA share a substantial number of links to \url{techcrunch.com}, a popular technology website; i.e., users who have tweeted at least once about these companies have an interest about subjects related to new technologies.
In fact, the top 10 list of 23andMe's set of tweets also include \url{lnkd.in}, \url{mashable.com}, and \url{entrepreneur.com}.
For AncestryDNA, we find \url{wn.nr}, another website related to contests and sweeps. %
There are thousands of tweets like \enquote{Enter for a chance to win a \$500 Gift Card! wn.nr/DRRrZq \#MemorialDaySweeps \#Entry}. 
We also note the presence of \url{woobox.com}, a marketing campaign website, responsible for organizing giveaways, as well as \url{giveaway.amazon.com}, an Amazon site organizing promotional sweepstakes.
\longVer{We believe this might be due to a large presence of bots, however, Botometer scores actually indicate these accounts are not.
Therefore, this behavior might be related to the fact that AncestryDNA, through their marketing campaigns, attract Twitter users who are generally active in looking for deals and sweeps.}
\shortVer{Botometer scores indicate these accounts are not actually bots, hence this might be related to the fact that AncestryDNA, through their marketing campaigns, attract Twitter users who are generally active in looking for deals and sweeps.}

\subsection{Discussion}
Overall, our user-level analysis shows that Twitter discourse related to genetic testing is dominated by US users and in general by those in English-speaking countries, but not necessarily by ``mainstream'' popular accounts.
However, 23andMe and AncestryDNA do regularly attract the attention of major news sites. 
Also, the majority of users involved in genetic testing discussion are not bots, so Twitter conversation is, to some extent, ``genuine.'' 
However, promotion and marketing campaigns end up attracting different kinds of users, and yield different levels of engagement.
Overall, we find that users interested in genetic testing appear less active than a random baseline, however, they are more likely to use Twitter to get information about their topics of interest, and in particular they are interested in technology and digital health subjects.

\longVer{

\section{Case Studies}\label{sec:use-case}
We take a closer look at ``negative'' tweets, following the sentiment analysis presented previously.
We also investigate the presence of users who post their genetic test results.

\subsection{Racism} 
We select all tweets with genetic testing keywords from users who yield a total sentiment score below -3, obtaining 3,605 tweets from 3,209 unique users. 
We then manually examine those with keywords 23andMe or AncestryDNA (1,725 and 167, respectively), and find several of them containing themes related to racism, hate, and privacy fears.

In particular, the ``ethnic'' breakdown provided by ancestry reports\footnote{E.g.~\url{https://permalinks.23andMe.com/pdf/samplereport_ancestrycomp.pdf}}
seems to spur several instances of negative-sentiment tweets associated with racism and disapproval of multi-cultural/multi-ethnic values.
For instance, a user with more than 3K followers self-describing as a \enquote{Yuge fan for Donald Trump}, tweets: \enquote{Get this race mixing shit off my time line!!} (Mar 23, 2017) %
in response to a 23andMe video about ancestry.
Another posts: \enquote{I wanna do that 23andMe so bad! I'm kinda scared what my results will be tho lmao I'm prob like half black tbh}%
(Jan 13, 2017), and gets a response: \enquote{I was too just do it and never tell anyone if you're a halfbreed haha}. 
Also, a user identifying as `American Fascist' tweets: \enquote{I'd like to get the @23andMe kit but, I'm worried about the results. Just my luck, I'd have non-white/kike ancestors. \#UltimateBlackpill} (May 30, 2017). %

Although we leave to future work an in-depth analysis of genetic testing related racism on the Web, we assess whether it may be systematic on Twitter, e.g., appearing also in tweets not scored as negative by SentiStrength. %
To this end, we search for the presence of hateful words using the \url{hatebase.org} dictionary, a crowdsourced list of 1K terms that indicate hate when referring to a third person, removing
words that are ambiguous or context-sensitive, as done by previous work~\cite{hine2017kek}.
Naturally, this is a best-effort approach since hateful terms might be used in non-hateful contexts (e.g., to refer to oneself), or, conversely, racist behavior can occur without hate words. Also, Twitter might be removing tweets with hate words as claimed in their hateful conduct policy.\footnote{\url{https://support.twitter.com/articles/20175050}}
Nonetheless, we do find instances of hate speech, e.g., anti-semitic tweets such as: \enquote{as long as there are khazar milkers to cause people to demand my 23andMe results, i will always be here to shitpost} (Nov 19, 2016),  %
or \enquote{@*** i would be pleased if you posted your 23andMe so i can confirm your khazar milkers are indeed genuine} (Dec 23, 2016).

Note that ``Khazar milkers'' refers to an anti-semitic theory on the origin of Jewish people from the 1900s~\cite{feldman2017khazar}
In a nutshell, it posits that Ashkenazi Jews are not descendant from Israelites, but from a tribe of Turkic origin that converted to Judaism.
23andMe issued ancestry reports that suggested Ashkenazi Jews in a given haplogroup were descendant from a single Khazarian ancestor.
Understanding the ancestry of Jewish people has been of interest to the genetics community for years, %
and the Khazar theory has been refuted repeatedly~\cite{behar2013khazar-abbrv}.
Nonetheless, the alt-right has exploited it to corroborate their anti-semitic beliefs~\cite{vice-alt-right}, and incorporate it into their collection of misleading/factually incorrect talking points.
In particular, ``khazar milkers'' was allegedly coined by the ``@***'' user mentioned above, and is used to imply a sort of succubus quality of Jewish women.

\subsection{Privacy} 
We also identify, among the most negative tweets, themes related to fears of privacy violation and data misuse.
Examples include \enquote{Is it me? Does the idea of \#23andMe seem a bit sinister? Do they keep the results? Who owns the results? Who owns 23andMe?}(Jan 1, 2016), %
\enquote{Same thing with 23andMe and similar companies. Indefinitely stored data with possible sinister future uses? \#blackmirror}(Nov 13, 2016), %
and \enquote{Why does this scare the hell out of me? How can our privacy ever be assured?} (Feb 27, 2016).
Searching for `privacy' and `private' in our keyword dataset  returns 1,991 tweets, mostly from 23andMe and Precision Medicine (1.1K and 625, resp.), which we proceed to examine both manually and from a temporal point of view (i.e., measuring daily volumes).
Overall, we find that privacy in the context of genetic testing appears to be a theme discussed recurrently on social media and a concern far from being addressed.
This is not entirely unexpected, considering that both the DTC market and the genomics landscape are evolving relatively fast, with regulation and understanding of data protection as well as informed consent often lagging behind, as also highlighted in prior work~\cite{mascalzoni2008informed,de2014genomic,phillips2016only}.

Interestingly, one of the peaks in tweets related to 23andMe and privacy
occurs on Oct 19, 2015 (with 152 tweets). %
As discussed in the Content Analysis section, this a relevant date w.r.t.~the FDA revoking their approval for 23andMe's health reports, which yields a peak in 23andMe tweets overall.
However, the FDA ruling had nothing to do with privacy, yet, it put 23andMe in the spotlight, possibly causing privacy concerns to resurface.
In fact, privacy and 23andMe discussions periodically appear in our dataset, even beyond tweets with negative sentiment, e.g., %
\enquote{I want to do \#23andMe but don't want a private company owning my genetic data. Anyone heard of any hacks to do it anonymously?} (Jul 13, 2017),
\enquote{@23andMe ur privacy policy describes how there is no privacy. How about u not share any data at all. I pay u and u send the results. Period} (Dec 8, 2015), %
\enquote{Should we be concerned about data collection and privacy with direct to consumer DNA testing companies like 23andMe?} (Apr 19, 2017).

\subsection{Users Sharing Test Results}

Finally, we investigate the presence of users who post their genetic test results,
aiming to estimate their number and shed light on their profiles. 
Given their popularity, we only do so for 23andMe and AncestryDNA.
Among other things, we believe this is important because health/ancestry reports may contain sensitive information about the individuals taking these tests, including their predisposition to diseases and their ethnic heritage~\cite{ayday2015whole}.

\descr{Methodology.} Finding all tweets that may include test results is quite hard, and arguably out of scope, thus, we focus on {\em screenshots} of genetic test results as we anecdotally find a non-negligible number in our dataset. 
These are actually almost exclusively ancestry results, even though 23andMe also provides health reports.
We start from the 4.5K/1.5K 23andMe/AncestryDNA tweets with images, but, since they are too many to be all manually examined, we use the following methodology. First, we build two sets with 100 ``ground truth'' images with screenshots of ancestry test results, one each for 23andMe and AncestryDNA, then, use Perceptual Hashing~\cite{phash2} to find similar images that are likely screenshots of test results too, and manually check them to exclude false positives.\footnote{\baselineskip=0.8\normalbaselineskip  Perceptual hash functions extract features from multimedia content and calculate hash values based on them. They can be used to compare two objects by calculating a distance/similarity score between two hash values; the objects are labeled as (perceptually) equal if the distance is below a chosen threshold~\cite{pHash}. We set the pHash distant threshold to 17 since it produces the best results in our setting.}

Overall, our approach likely yields a conservative estimate, nonetheless, it constitutes a best-effort approach to identify and analyze tweets including test results.
We obtain 366 and 204 images for, respectively, 23andMe and AncestryDNA. Upon manual examination, we find and remove 58 (16\%) and 26 (13\%) false positives.
Thus, we estimate a lower bound of 0.23\% and 0.60\% of 23andMe and AncestryDNA tweets containing ancestry test screenshots (and 3.40\% and 5.15\% of tweets with pictures).

\descr{Tweets content.} We manually examine the 486 tweets with screenshots of ancestry test results, finding that users often appear to be somewhat enthusiastic about their experience.
In some cases, we note a feeling of ``relief'', sometimes expressed in a humorous way, when the results show they are predominantly  ``white'':  about 10\% of tweets with screenshots include the word white. Examples include: \enquote{23andMe confirms: I'm super white.}, \enquote{Got my @23andMe results back today. I'm super white. Like, rice on a paper plate with a glass of milk in a snowstorm, white.}

\descr{User Analysis.} We also crawl the last 1,000 tweets of the 308 users who have posted screenshots with test results, and analyze them as done for the random sample discussed in Sec.~\ref{sec:user-analysis}. 
We find that their most commonly used hashtags are indeed related to genetic testing, confirming that the users who do genetic tests are actually generally interested in the subject.
It is also interesting to find \#maga to be the second most common hashtag for users who post 23andMe results (appearing in 431 tweets). 
Then, looking at the top domains shared by these users, we do not observe surprising difference compared to those from a random sample users (see Table~\ref{tab:top_domains_user}), although we find more social media services like Instagram and Facebook.
}

\section{Conclusion}\label{sec:discussion}

This paper presented a large-scale analysis of Twitter discourse related to genetic testing.\longVer{ We examined more than 300K tweets containing 13 relevant keywords as well as 12M tweets posted by more than 100K accounts that have shown interest in genetic testing.}
We found that the discourse related to genetic testing is often influenced by news and technology websites, and by a group of tech-savvy users who are overall interested in tech and digital health.
Overall, users tweeting about genetic testing are mostly in the US and other English-speaking countries, while we do not find evidence of extensive influence of social bots. 
However, the broad conversation seems to be dominated by users that might have a vested interest in its success, e.g., specialist journalists, medical professionals, entrepreneurs, etc.
This is \longVer{particularly }evident in the tweets related to genomics initiatives, which are mainly discussed by highly engaged users and which are influenced, at least in terms of volume, by important announcements. 
Moreover, we noticed that the two most popular DTC companies, 23andMe and AncestryDNA, also generate the most tweets, however, although 23andMe has half the customers, it produces almost 5 times more tweets, also due to controversy around their failure to get FDA approval in 2015. 
We also observed a clear distinction in the marketing efforts \longVer{undertaken }by different companies, which ends up influencing users' engagement on Twitter.

\longVer{Our work is particularly timely as genetic testing and genomics initiatives are increasingly often associated to ethical, legal, and societal concerns~\cite{ftc2017dna}.
In this context, our analysis sheds light not only on who tweets about genetic testing, what they talk about, and how they use Twitter, but also on groups utilizing genetic testing to push racist agendas and users expressing privacy concerns.
We also found a number of enthusiastic users who broadcast their test results through screenshots notwithstanding possible privacy implications. O}%
\shortVer{In the full version of the paper~\cite{full}, we also shed light on tweets including racism, privacy and data misuse fears, and look for instances of users sharing screenshots of their test results; o}%
verall, our findings motivate future work studying other social media platforms and health forums/websites,
as well as for in-depth qualitative studies of the relation between genetic testing and racism and privacy fears on social media.

\small
\bibliographystyle{abbrv}
%\bibliography{bibliography}

\end{document}